\documentclass[%
 aip,
 amsmath,amssymb,
 reprint,%
]{revtex4-1}

\usepackage{graphicx}
\usepackage{dcolumn}
\usepackage{bm}

\usepackage[utf8]{inputenc}
\usepackage[T1]{fontenc}
\usepackage{mathptmx}
\usepackage{etoolbox}

\usepackage{float}
\usepackage{subfigure}
\usepackage{overpic}
\usepackage{caption}
\usepackage{textcomp}
\usepackage{gensymb}
\usepackage{siunitx}

\makeatletter
\def\@email#1#2{%
 \endgroup
 \patchcmd{\titleblock@produce}
  {\frontmatter@RRAPformat}
  {\frontmatter@RRAPformat{\produce@RRAP{*#1\href{mailto:#2}{#2}}}\frontmatter@RRAPformat}
  {}{}
}%
\makeatother
\begin{document}

\preprint{AIP/123-QED}

\title{First-principles analysis of the warm dense plasma jets in Double-Cone Ignition experiments}
\author{N.-Y. Shi}
 
\author{J.-H. Liang}%
 \homepage{liangjionghang@sjtu.edu.cn}
\affiliation{ 
National Key Laboratory for Dark Matter, Key Laboratory for Laser Plasmas, Department of Physics and Astronomy, and Collaborative Innovation Center of IFSA, Shanghai Jiao Tong University, Shanghai 200240, People’s Republic of China
}%

\author{C.-J. Mo}
\affiliation{%
Beijing Computational Science Research Center, Beijing 100193, China
}%
\author{D. Wu}
 \homepage{dwu.phys@sjtu.edu.cn}

\author{X.-H. Yuan}
\affiliation{ 
National Key Laboratory for Dark Matter, Key Laboratory for Laser Plasmas, Department of Physics and Astronomy, and Collaborative Innovation Center of IFSA, Shanghai Jiao Tong University, Shanghai 200240, People’s Republic of China
}%
\author{J. Zhang}
 \homepage{jzhang@iphy.ac.cn}
\affiliation{ 
National Key Laboratory for Dark Matter, Key Laboratory for Laser Plasmas, Department of Physics and Astronomy, and Collaborative Innovation Center of IFSA, Shanghai Jiao Tong University, Shanghai 200240, People’s Republic of China
}%
\affiliation{Institute of Physics, Chinese Academy of Sciences, Beijing 100190, People’s Republic of China}
\date{\today}

\begin{abstract}
Double-Cone Ignition (DCI) differs from the traditional laser-driven inertial confinement fusions by relying on gold cones for transverse filtering to achieve warm dense plasma, thereby reducing the energy required during the compression process. Thus, the state of the plasma ejected from the gold cones directly reflects the energy conversion efficiency and influences the subsequent fusion process. In this paper, we analyze the x-ray Thomson scattering data from the earlier stage DCI experiments \cite{zhang2020double}. We combine the imaginary-time correlation function method with first-principles methods to decouple the diagnosis of temperature and density and obtain the temperature and density diagnostic outputs: $25$ $\mathrm{eV}$ and $8\pm2$ $\mathrm{g/cc}$. In the analysis, we consider the effect of multi-element mixing and determine the gold impurity ratio to be $0.162 \pm 0.015 \%$ based on the experimental spectrum. These detailed analysis results demonstrate the role of the gold cone in achieving plasma compression and confirm that the gold impurities are within an acceptable range, providing valuable references for future experiments.
\end{abstract}

\maketitle

\section{\label{sec:level1}Introduction}

Based on previous research on direct and indirect drive central ignition scheme, the Double-Cone Ignition scheme (DCI) is employed as an innovative laser inertial confinement ignition scheme and is expected to reduce the energy demand for compression and heating \cite{zhang2020double}. Under the influence of high-power lasers, the deuterium-tritium (D-T) fuel, placed within two concentric gold cones, is accelerated into a supersonic jet along the axis of the cones. The two jets meet and collide to form a high-density plasma, and high-energy fast electrons are injected to heat the plasma. In this scheme, the plasma at the cone exit, due to synchronized compression and preheating \cite{zhang2020double}, is a typical warm dense matter (WDM) \cite{unknown}. For DCI scheme, the state of the WDM plasma ejected from the gold cones serves as a direct indicator of energy conversion efficiency and impacts the subsequent fusion process.

WDM, with temperatures of several $\mathrm{eV}$ and densities close to solid states, is broadly concerned in inertial confinement fusion, planetary physics and laboratory astrophysics \cite{Lahmann_2023}. Although WDM can be prepared in laboratories under extreme conditions \cite{graziani2014frontiers,SAUNDERS201886}, experiments on WDM still face many challenges, such as the generation and maintenance of extreme states \cite{Li2022}, material interface effects and contamination \cite{PhysRevE.96.063310}, and the diagnostics of transient processes \cite{Mahieu2018}, making it difficult to repeat high-precision experiments. Among diagnostic techniques, X-ray Thomson scattering (XRTS) has become a promising and reliable method to diagnose the internal properties of WDM \cite{graziani2014frontiers,10.1063/1.4890215,Lahmann_2023}.

Following the proposal of the Chihara model \cite{10.1063/1.4985729}, XRTS diagnostics have evolved over the years into a mature diagnostic technique. The reviewer paper of S. H. Glenzer and R. Redmer \cite{RevModPhys.81.1625} describes multiple aspects of theory and experiment in detail, providing a foundation for researchers; the diagnostics of aluminum by P. Sperling et al. \cite{PhysRevLett.115.115001}, beryllium by A.L. Kritcher et al. \cite{PhysRevLett.107.015002}, and CH by L.B. Fletcher et al. \cite{PhysRevLett.112.145004} demonstrate the applicability of XRTS diagnostics to different elements and experimental conditions. Compared to the Random Phase Approximation method used in early Chihara models, first-principles methods \cite{PhysRevE.92.013103,PhysRevLett.120.205002} offer a more accurate interpretation of XRTS spectra. Recently, T. Dornheim et al. \cite{Dornheim2022} proposed a new model-free approach for diagnosing XRTS temperature. Building on previous research, this article will focus on the special experimental environment of DCI and present a more accurate analysis of the previous experiment\cite{Zhang2022}.

\begin{figure*}[htbp]
	\centering
    \begin{minipage}[b]{.49\linewidth}
        \centering
            \begin{overpic}[width=0.95\linewidth]{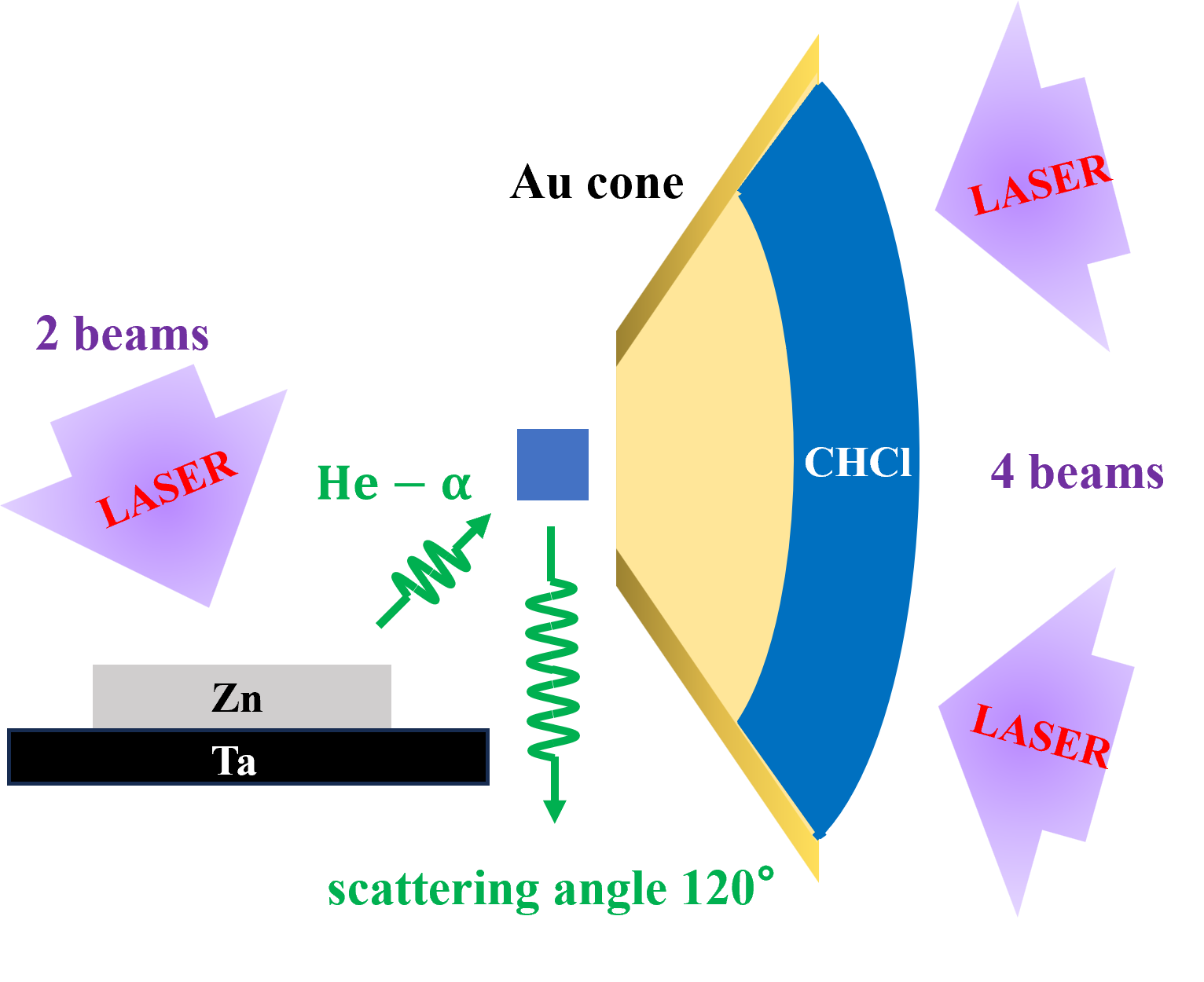}
            \put(0,75){\large\textbf{(a)}}
        \end{overpic}
    \end{minipage}
    \begin{minipage}[b]{.49\linewidth}
        \centering
           \begin{overpic}[width=0.95\linewidth]{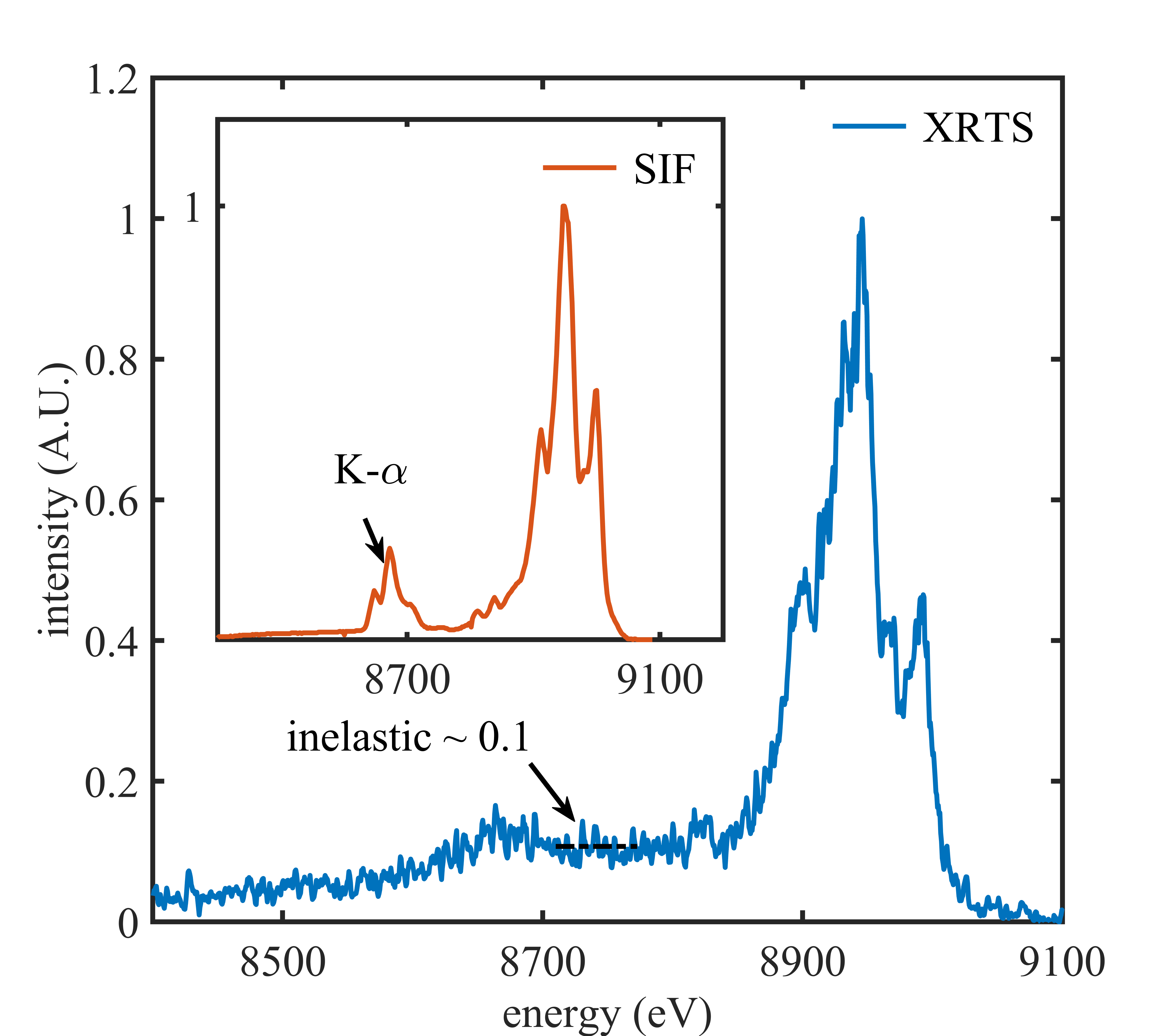}
           \put(0,75){\large\textbf{(b)}}
        \end{overpic}
    \end{minipage}
    \caption{Experiment setup and result: (a) Experiment setup: four beams driven laser compressed the CHCl shell through the gold cone and the X-ray targeted the plasma at the cone exit; (b) Display of XRTS spectrum (blue) and SIF spectrum (red): black dotted line shows the inelastic peak, which is around 0.1. The left peak in the SIF spectrum is K-$\alpha$ line.}
    \label{experiment}
\end{figure*}

The experiment was carried out at the Shenguang-$\mathrm{\uppercase\expandafter{\romannumeral2}}$ upgrade facility (SG-$\mathrm{\uppercase\expandafter{\romannumeral2}}$ UP), which carries eight nanosecond lasers and one picosecond laser. Each nanosecond laser has a third-harmonic output capability of 3 $\mathrm{kJ}$, while the picosecond laser can output 1000 $\mathrm{J}$ with a pulse width of 10 $\mathrm{ps}$. The hollow cone used for compression in the experiment was made of gold, with a wall thickness of 20 $\mathrm{\mu m}$. The angle of the gold cone was $100^{\circ}$, which corresponds to the laser arrangement. The front end of the gold cone had an aperture of 100 $\mathrm{\mu m}$. The spherical shell used for compression was made of C16H14Cl2, with an inner radius of 450 $\mathrm{\mu m}$ and a thickness of 45 $\mathrm{\mu m}$. The X-ray source for Thomson scattering was generated by two nanosecond lasers with a total energy of 3.5 $\mathrm{kJ}$, driving a 10 $\mathrm{\mu m}$ thick Zn target to produce He-$\alpha$ (8.95 $\mathrm{keV}$) X-rays. The scattering angle selected in the experiment was $120^{\circ}$, and a von Hamos configured HOPG (highly oriented pyrolytic graphite) crystal spectrometer was used for spectral measurement. The crystal had a radius of curvature of 170 mm, dimensions of 30 $\mathrm{mm}$ $\times$ 70 $\mathrm{mm}$, and a mosaic angle of $0.4^{\circ}$. The measurement range of the spectrometer ranged from $8.4$ to $9.1$ $\mathrm{keV}$, with a spectral resolution of 12.9 $\mathrm{eV}$, fully meeting the measurement requirements. The experiment setup is shown in Fig. \ref{experiment}(a) and corresponding XRTS result is shown in  Fig. \ref{experiment}(b).

Previous XRTS diagnostics \cite{PhysRevE.67.026412} were based on relatively pure and single target materials, generally containing only one to two main elements. However, the XRTS spectrum to be diagnosed in the experiment has a target pellet composed of C, H, and Cl in a fixed ratio, with an unknown proportion of high-Z element gold mixed in. It is necessary to accurately calculate the XRTS spectral characteristics at different temperatures and densities while also being able to distinguish the ratios of the four elements, which poses extremely high demands on the diagnostic model. This article has essentially solved this difficult problem through first-principles methods and the powerful computational capabilities of DFT, expanding XRTS from temperature-density diagnostics to multi-element impurity diagnostics, making an important contribution to understanding the current status of DCI experiments.

This article is organized as follows. First, we diagnose the WDM plasma temperature as $25$ $\mathrm{eV}$ based on the ITCF method, which serves as a basis for the subsequent diagnosis. Next, this paper employs first-principles methods to calculate the non-elastic spectrum at different densities, diagnosing the electron density as $1.7\times10^{24}$ $\mathrm{cm^{-3}}$, with equivalent density of CHCl to be $8\pm2$ $\mathrm{g/cc}$. Finally, by comparing the differences in elastic peak heights between experiments and simulations, we determine the gold impurity ratio to be $0.162 \pm 0.002 \mathrm{\%}$.

\section{\label{sec:level1}first-principles method}
To begin with theoretically, the spectrum intensity is proportional to the particle number N and the electron dynamic structure factor $S(\mathbf{q},\omega)$, with the latter well-explained by Chihara model \cite{10.1063/1.4985729}:
\begin{align}
\begin{split}
   S(\mathbf{q},\omega)=S_{ii}(\mathbf{q},\omega)+Z_fS_{ee}(\mathbf{q},\omega)+S_{bf}(\mathbf{q},\omega)
\label{1}
\end{split}
\end{align}
In this model, the first bound-to-bound term represents quasi-elastic Rayleigh scattering, which includes the effects of bound and free electrons on ion behavior. The second free-to-free term involves scattering contribution from free electrons that do not follow ion motion, and the third bound-to-free term is inelastic scattering by strongly bound core electrons, which arises from Raman transitions to the continuum of core electrons within an ion. Based on the model, we employ a first-principles method based on density functional theory (DFT) to calculate each part of the Chihara model, providing accurate XRTS numerical simulation results.

\textit{Bound-to-Bound} The quasi-elastic Rayleigh scattering in Eq. (\ref{1}) is written as \cite{PhysRevB.102.195127}:
\begin{align}
\begin{split}
   S_{ii}(\mathbf{q},\omega)=\langle \vert \sum_n f_n\langle n\vert e^{-i\mathbf{q}\cdot\mathbf{r}} \vert n \rangle\vert^2\rangle \delta(\omega)+\\\langle \vert \sum_n f_n(1-f_n)\vert\langle n\vert e^{-i\mathbf{q}\cdot\mathbf{r}} \vert n \rangle \vert ^2 \rangle \delta(\omega)
\label{2}
\end{split}
\end{align}
where $f_n=1/[e^{(\varepsilon_n-\mu)/T}+1]$ is the Fermi-Dirac occupation of the $n$th level, and $\langle n\vert e^{-i\mathbf{q}\cdot\mathbf{r}} \vert m \rangle =\int d\mathbf{r} e^{-i\mathbf{q}\cdot\mathbf{r}}\phi_n^*(\mathbf{r})\phi_m(\mathbf{r})$ with $\mathbf{r}$ the coordinates of the particle and $\phi_m(\mathbf{r})$ the $m$th orbital wave function. The $\langle...\rangle$ represents the average of ionic configurations. Since the calculation reveals that the second term of Eq. (\ref{1}) has a relatively minor contribution, we can focus on the first term and further write it as \cite{PhysRevB.102.195127}:
\begin{align}
\begin{split}
   S_{ii}(\mathbf{q},\omega)= N_i\vert N(\mathbf{q})\vert^2S_{ii}(\mathbf{q})\delta(\omega)
\label{3}
\end{split}
\end{align}
where $N_i$ is the number of ions, $N(\mathbf{q})$ is the electronic form factor and $S_{ii}(\mathbf{q})$ is the static ionic structure factor.

The above formulas are for one-component plasma, while in our experiment and simulation, the plasma is of multiple components. Therefore, Eq. (\ref{3}) is changed into
\begin{align}
\begin{split}
   S_{ii}(\mathbf{q},\omega)= N_i\sum_{\mu\nu}\sqrt{x_{\mu}x_{\nu}}N_{\mu}(\mathbf{q})N_{\nu}(\mathbf{q})S_{\mu\nu}(\mathbf{q})\delta(\omega)
\label{4}
\end{split}
\end{align}
where $\mu$,$\nu$ represents components such as C and H, $x_j$ $(j=\mu,\nu)$ means the percentage of the j-component in the whole system.

\textit{Free-to-Free} The inelastic scattering from free electrons in Eq. (\ref{1}) can be calculated from the electronic density response function $\chi_{ee}(\mathbf{q},\omega)$ through the fluctuation-dissipation theorem as
\begin{align}
\begin{split}
   S_{ee}(\mathbf{q},\omega)=(\frac{-1}{1-\exp(-\omega/T_e)})\mathrm{Im}(\frac{\chi_{ee}(\mathbf{q},\omega)}{\pi n_e})
\label{5}
\end{split}
\end{align}
where $n_e$ is the electronic density \cite{PhysRevLett.120.205002} .

The electronic wave functions are firstly calculated by the Mermin’s finite temperature version of the DFT, then the $\chi_{ee}(\mathbf{q},\omega)$ and $S_{ee}(\mathbf{q},\omega)$ can be calculated using time-dependent density functional theory (TDDFT) with a linear response perturbation formula. The DFT calculations are carried out using the QUANTUM ESPRESSO package, and the $\chi_{ee}(\mathbf{q},\omega)$ and $S_{ee}(\mathbf{q},\omega)$ are calculated using YAMBO code \cite{PhysRevLett.120.205002}.

\textit{Bound-to-Free} Impulse approximation (IA) is selected to calculated the bound-free transition in Eq. (\ref{1}), as it is more accurate in the high-energy region compared with other methods: hydrogenic model, plane-wave form factor approximation, etc \cite{10.1063/1.4790659}. In the limit of large energy transfer $\omega$ relative to the initial-state binding energy $E_B$, the spectrum is determined by the initial-state electronic momentum distribution. By inserting a complete set of momentum eigenstates, $S_{bf}(\mathbf{q},\omega)$ changes into
\begin{align}
\begin{split}
   S_{bf}(\mathbf{q},\omega)=N_i\sum_{i\in B_s}g_i f_i S_i(\mathbf{q},\omega),\\S_i(\mathbf{q},\omega)=\frac{2\pi}{\lvert \mathbf{q} \rvert}\int_{|\omega/\lvert \mathbf{q} \rvert-\lvert \mathbf{q} \rvert/2|}^{\infty}\mathbf{p}\rho_i(\mathbf{p})d\mathbf{p}
\label{6}
\end{split}
\end{align}
where $N_i$ is atom number, $g_i$ represents degenerancy, $f_i$ stands for Fermi-Dirac occupation, $B_s$ represents orbitals for bound electrons and $\rho_i(p)=(2\pi)^{-3}|\langle i|p\rangle|^2$ is the initial-state momentum density.

\section{\label{sec:level1}Results and Discussions}
In this work, we diagnose the XRTS experimental data based on the characteristics of multi-element and high-Z element impurity in DCI. The diagnostic procedure comprises three sequential components: (1) Temperature diagnosis employing the imaginary-time correlation function (ITCF) method, which establishes the fundamental basis for subsequent analyses; (2) Density diagnosis conducted via first-principles calculations, whose results provide critical insights into gold impurity effects; and (3) Gold impurity diagnosis based on first-principles approaches, where judicious approximations are implemented to obtain reliable estimates of gold impurity ratio.

\subsection{\label{sec:level2}Temperature Diagnosis}
Previous XRTS diagnostic methods obtain temperature and density diagnostic information by fitting the peak height and broadening of the inelastic peak with simulated curves. This method will inevitably require traversing a large number of temperature-density points to obtain the best fitting results. However, because of the long calculation time of the TDDFT method, traversing temperature-density points is neither economical nor rapid. Therefore, we use the ITCF method \cite{Dornheim2022,10.1063/5.0222072} to decouple the temperature-density coupling in curve fitting, which is robust with respect to the background noise in experiment \cite{10.1063/5.0222072,Dornheim2022}.

\begin{figure}[htbp]
\centering
\includegraphics[scale=0.5]{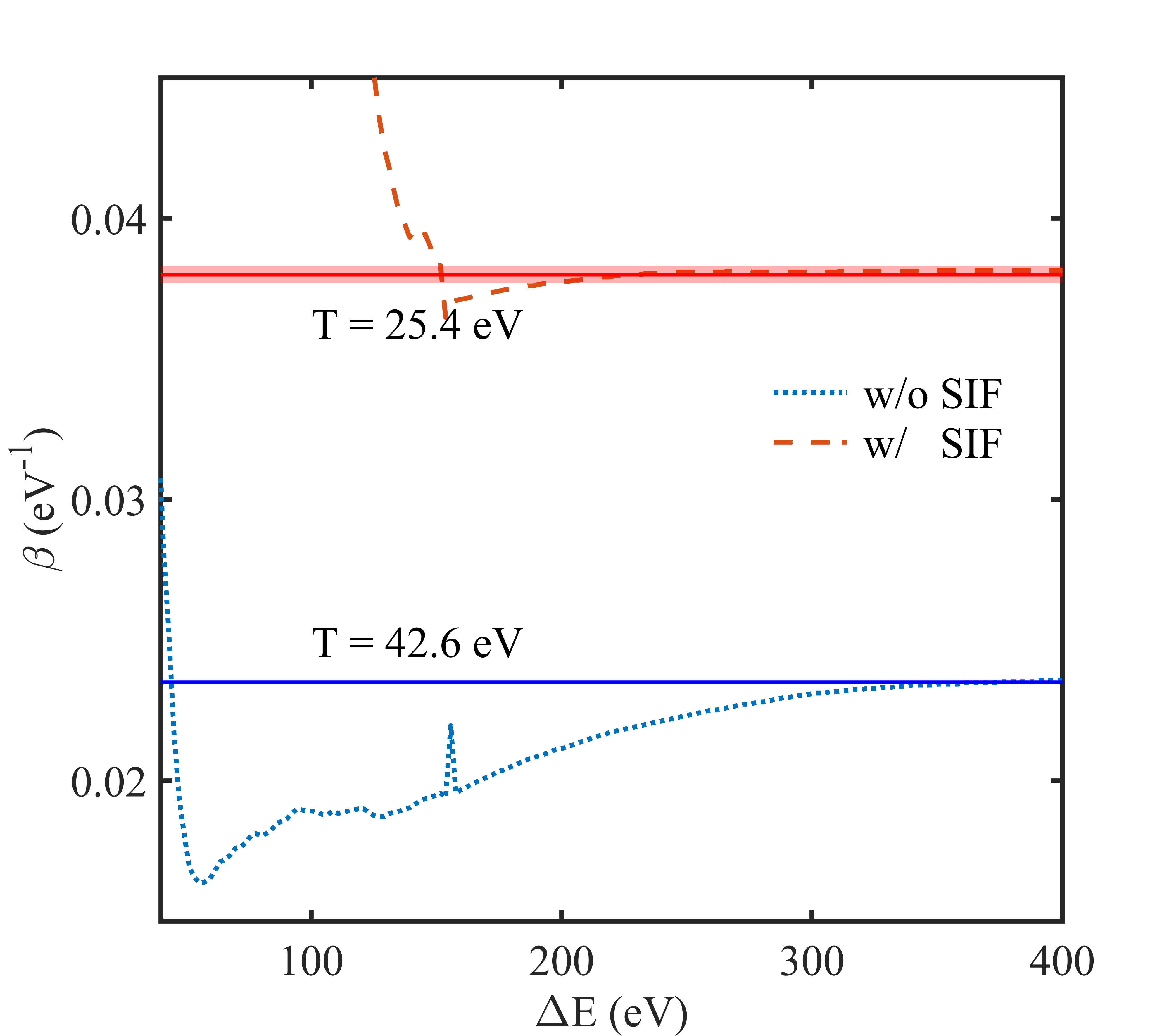}
\caption{Temperature diagnosis: convergence of model-free diagnosis with respect to the integration boundary E of $\mathcal{L}[S(\mathbf{q},E)]$ with correcting for SIF (red) and without (blue).}
\label{beta}
\end{figure}

Fig. \ref{beta} is the temperature diagnosis of the XRTS spectrum shown in Fig. \ref{experiment}. The blue line is the XRTS spectrum and the red one is the corresponding source and instrument function (SIF). Through expansion of the integration boundary, the minimum values among successive integration results are identified to determine the parameter $\beta$, shown in Fig. \ref{beta}. The red dots shows the convergence of $\beta$ with respect to the integration boundary $\Delta E$, and the temperature extracted from the experiment data is $T=25.4\pm 0.3$ $\mathrm{eV}$. We also calculate the convergence without considering SIF, shown as blue line, and the extracted temperature is $T=42.6\pm 0.1$ $\mathrm{eV}$.

It has to be pointed out that the XRTS and SIF spectra are not obtained in the same round of experiment, so even if the experiment setup is the same except the existence of the target in the cone, the interaction time between the laser and the Zn target, and the effective scattering time between the X-rays and the jet plasma may differ, leading to discrepancies between the obtained SIF and the actual SIF spectrum. This is shown in the broader half-width and higher $K$-$\alpha$ peak of SIF. Furthermore, the spectral range is insufficient to extend beyond 9100 eV. However, given that the spectral intensity at 9100 eV approaches zero, we conclude that contributions from energies above this threshold are negligible in the scattering spectrum. Nevertheless, the ITCF method provides a reasonable and meaningful diagnostic result, so we take it as the final temperature diagnostic result and diagnose the density based on this. 

\subsection{\label{sec:level2}Density Diagnosis}
With the diagnosed temperature of $25$ $\mathrm{eV}$, we start to diagnose the XRTS spectrum through curve fitting. In the experiment, the target consists of 16 C, 14 H and 2 Cl, so 8 C, 7 H and 1 Cl ions are randomly put in the system and  the lattice parameter is adjusted to change the density. We select molten CHCl in random structures with $T_i=T_e$, therefore, the ionic configurations are generated with a $\Gamma$-point first-principle molecular dynamics (FPMD) simulation at given temperature $T_e=25$ $\mathrm{eV}$. Orbitals are generated using the FT-DFT calculation with the local density approximation (LDA) and a $6\times6\times6$ shifted k-point mesh is used to resolve the Brillouin zone. An energy cutoff of 250 Ry and 1500 energy bands are used to ensure the convergence of wave functions. The effective ion-electron interaction is represented by three normal conserving pseudopotentials, one for C with four valence electrons, another for H with one valence electron and the other for Cl with seven valence electrons. Therefore, inner orbits of C and Cl are considered in the calculation of $S_{bf}(\mathbf{q},\omega)$ in Eq. (\ref{6}).

In the calculation of $\chi_{ee}(\mathbf{q},\omega)$ and $S_{ee}(\mathbf{q},\omega)$ in Eq. (\ref{5}) with the TDDFT method, the resolution of $\omega$ is $1$ $\mathrm{eV}$, and the adiabatic local density approximation is adopted. The transferred momentum is set to be $\lvert \mathbf{q} \rvert =4.16$ $\mathrm{bohr}^{-1}$, corresponding to the $120^{\circ}$ scattering angle and the He-$\alpha$ line of Zn at $8950$ $\mathrm{eV}$ in the experiment.

In the calculation of $N(\mathbf{q})$ and $S_{ii}(\mathbf{q})$ in Eq. (\ref{3}), 16 C, 14 H and 2 Cl ions are randomly put in the system and the lattice parameter is adjusted to change the density. Molten CHCl in random structures with $T_i=T_e=25$ $\mathrm{eV}$ is selected and ionic configurations are generated with a $\Gamma$-point first-principle molecular dynamics (FPMD) simulation with an energy cutoff of 130 Ry and 700 energy bands. The Brillouin zone is sampled with the $\Gamma$-point and the time step is set to $0.05$ $\mathrm{fs}$ due to the high temperature. After equilibrium is reached, $\sim 30000$ configurations are saved for the calculation of $S_{ii}(\mathbf{q})$. With Eq. (\ref{4}), $S_{\mu\nu}(\mathbf{q})$ representing various mixtures of components are calculated with all directions of $\mathbf{q}$ averaged, and the corresponding $N_{\mu}(\mathbf{q})$ and $N_{\nu}(\mathbf{q})$ are also obtained.
\begin{figure}[htbp]
\centering
\includegraphics[scale=0.49]{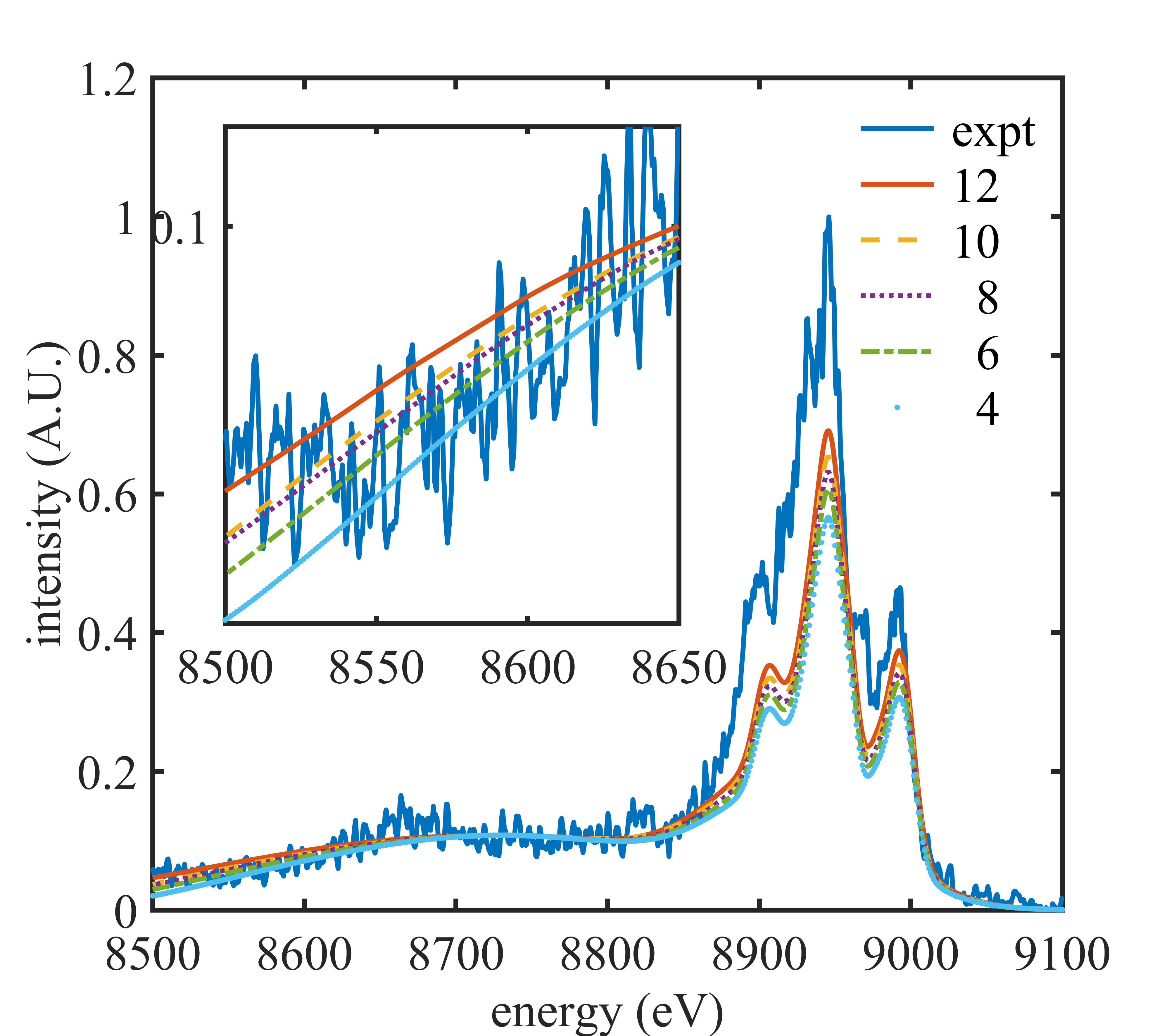}
\caption{Density diagnosis: simulations of various densities: $12$ $\mathrm{g/cc}$ (red), $10$ $\mathrm{g/cc}$ (yellow), $8$ $\mathrm{g/cc}$ (purple), $6$ $\mathrm{g/cc}$ (green) and $4$ $\mathrm{g/cc}$ (blue dot), together with the experiment data (blue). The inelastic part is magnified and displayed in the inset at the top left. By examining the fit of the curve, it is determined that the best density fit result is $8\pm2$ $\mathrm{g/cc}$.}
\label{density}
\end{figure}

We combine all parts of the XRTS simulation and obtain the spectra corresponding to different densities shown in Fig. \ref{density}. All the spectra are convoluted with SIF to take into account the probe's bandwidth. The inelastic part is magnified and displayed in the inset at the top left. Upon examination of the magnified inelastic scattering curve fit, it is determined that the best density fit result is $8\pm2$ $\mathrm{g/cc}$ since $12$ $\mathrm{g/cc}$ (red) is higher and $4$ $\mathrm{g/cc}$ (blue dot) is lower in the $<8550$ region. Uncertainty results from experimental noise in the data, such as the $K$-$\alpha$ peak around $8670$ $\mathrm{eV}$ and the unexplained long tail in the $8400$-$8500$ $\mathrm{eV}$ region. In the fitting, we can observe a significant discrepancy between theory and experiment in the elastic scattering part. This phenomenon inspires us to investigate the impact of gold impurity on XRTS diagnostics.

\begin{figure*}[htbp]
	\centering
    \begin{minipage}[b]{.49\linewidth}
        \centering
            \begin{overpic}[width=0.95\linewidth]{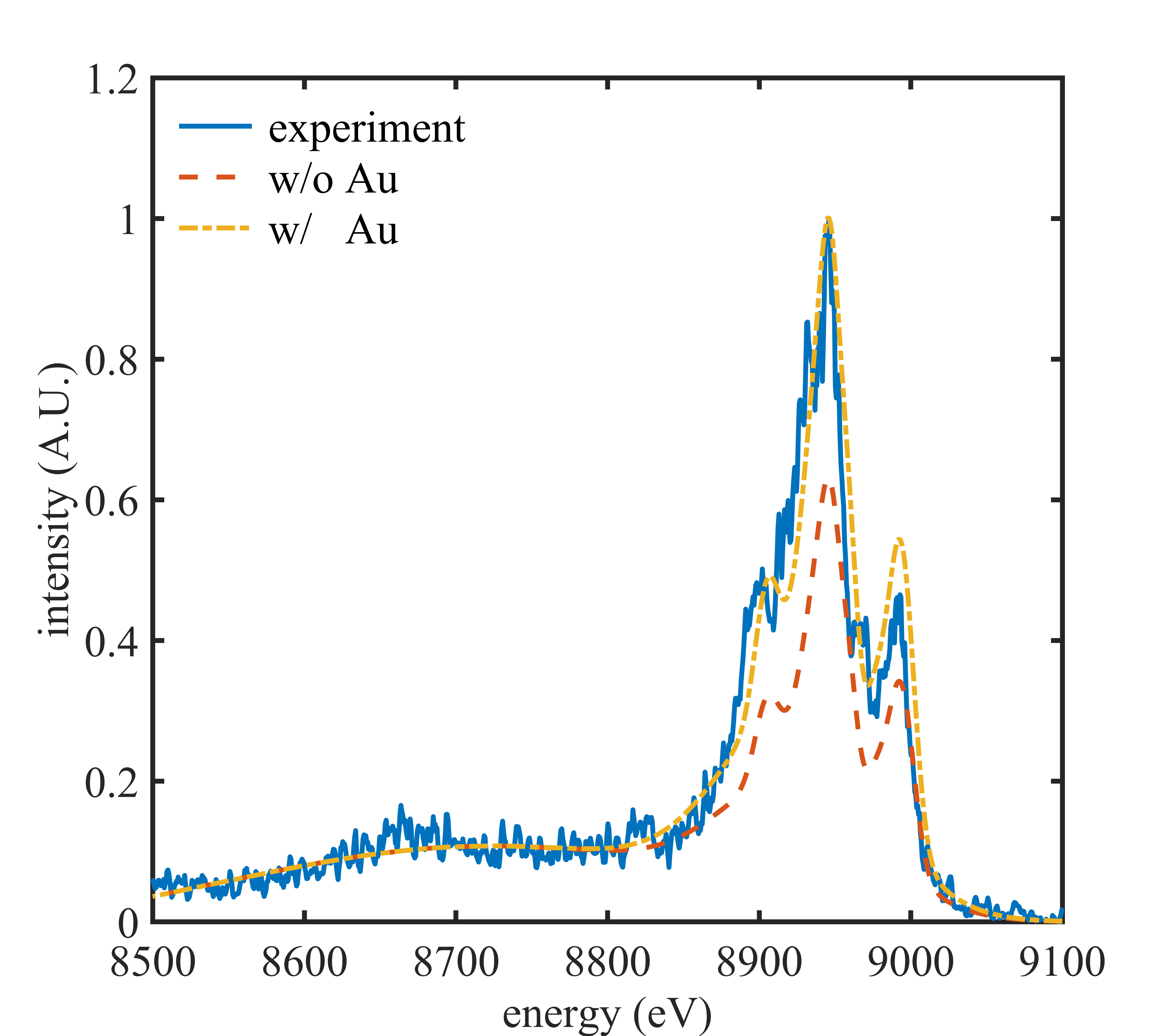}
            \put(0,77){\large\textbf{(a)}}
        \end{overpic}
    \end{minipage}
    \begin{minipage}[b]{.49\linewidth}
        \centering
           \begin{overpic}[width=0.95\linewidth]{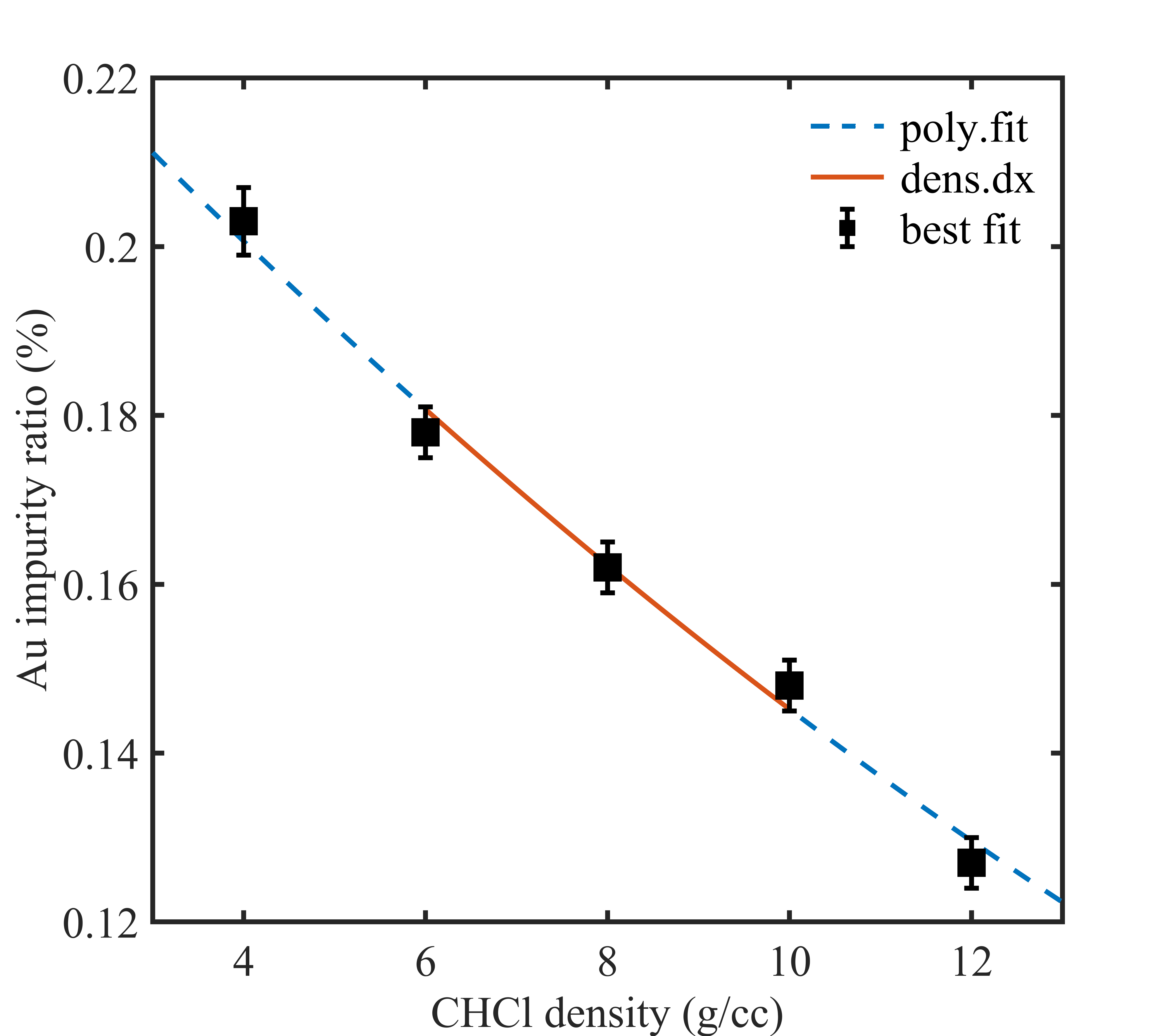}
           \put(0,77){\large\textbf{(b)}}
        \end{overpic}
    \end{minipage}
    \caption{Gold impurity diagnosis: (a) The fitting of the experiment data (blue), with temperature $25$ $\mathrm{eV}$ and electron density $1.7\times10^{24}$ $\mathrm{cm^{-3}}$, i.e. $8$ $\mathrm{g/cc}$, includes simulation with gold (yellow) and without gold (red). The corresponding gold impurity rate is $0.162 \pm 0.002\%$; (b) The black solid dots shows the doping rate of Au with respect to the density. The blue dotted line shows the general trend of the rate as a function of density and the red solid line represents the diagnosed density range.}
    \label{au}
\end{figure*}

\subsection{\label{sec:level2}Gold Impurity Diagnosis}
The inelastic peak profile is predominantly determined by temperature and electron density. The minor effects induced by gold doping can be neglected during diagnostic procedures, as detailed in Appendix B. The substantial number of inner-shell electrons in gold can generate intense bound-free transitions. However, given their negligible concentration, the resulting transition intensity remains lower than that originating from carbon atoms. Computational analyses have previously established that carbon-bound-free transitions do not contribute significantly to the overall spectrum. Consequently, gold-bound-free transitions can likewise be disregarded in our calculations. Ultimately, the most significant effect of gold impurity is reflected in the elastic peak. The large number of tightly bound inner electrons makes the electronic form factor $N(\mathbf{q})$ of gold a high value, which can significantly enhance the intensity of the elastic peak.

Because of the extremely low impurity ratio of Au, when the ionic configuration is precisely calculated through DFT-MD, it is necessary to expand the system to hundreds of atoms. This not only occupies a large amount of computational resources and takes a long time, but may also generate a large amount of useless data due to the unknown ratio and may even be incalculable. Therefore, when calculating $S_{\mu\nu}(\mathbf{q})$, we use the following approximation: Past calculations prove that under back-scattering condition, in $S_{\mu\nu}(\mathbf{q})$, all the $\mu$-$\mu$ components are close to 1, and all the $\mu$-$\nu$ components are close to 0, shown in Appendix C. Therefore, we take the $\mu$-$\mu$ components as 1, and the $\mu$-$\nu$ components as 0. At the same time, fix the ratio of CHCl to configuration of C16H14Cl2, which allows us to rewrite Eq. (\ref{4}) as 
\begin{align}
\begin{split}
   S_{ii}(\mathbf{q})=\frac{16}{32+x}N_{C}^2(\mathbf{q})+\frac{14}{32+x}N_{H}^2(\mathbf{q})\\+\frac{2}{32+x}N_{Cl}^2(\mathbf{q})+\frac{x}{32+x}N_{Au}^2(\mathbf{q})
\end{split}
\end{align}
where x represents the particle number of Au doping in the system.

Fig. \ref{au}(a) illustrates the impact of gold impurity on XRTS diagnostics under a density condition of 8 $\mathrm{g/cc}$. The difference in the height of the elastic peak between the yellow curve (with gold impurity) and the red curve (without gold impurity) intuitively shows the effect of gold. At the same time, the yellow curve and the blue experiment data form a good fit despite the difference resulting from SIF, which is convoluted in the simulation.

Fig. \ref{au}(b) shows the relationship between the gold impurity rate and the density in the simulation. Each black dot corresponds to the density for which simulations were performed in this paper, and the best-fit density of $8\pm2$ $\mathrm{g/cc}$ has a gold impurity rate $0.162 \pm 0.015\%$. It can be observed that the gold doping ratio decreases with increasing density. The blue line is a polynomial fit to the black dots, indicating only the general trend. The slight impurity ratio is consistent with the reasons for our approximations in the previous text. It also reflects a common phenomenon of high-Z elements impurity in XRTS: a small proportion can significantly impact the spectrum, especially the elastic peak. By listing the relationship between the gold impurity rate and density, we can provide a sufficiently accurate and reasonable description of the gold impurity situation in this experiment.

\section{\label{sec:level1}Conclusions} 

Building on previous research, we diagnose the XRTS experimental data based on the characteristics of multi-element and high-Z element impurity in DCI. First, we use the ITCF method to conveniently and directly obtain the temperature at $25$ $\mathrm{eV}$, which serves as the basis for subsequent diagnosis. Second, we employ first-principles methods to calculate the complete spectrum at different densities, diagnosing the density as $8\pm2$ $\mathrm{g/cc}$, and reflecting a high-Z element impurity through fitting. Finally, we innovatively include gold in the calculation scope, obtaining the best-fitted gold impurity rate at $0.162 \pm 0.015 \%$ and illustrating the relationship between gold impurity rate and density through impact on the ratio of elastic peak to inelastic peak.

In this work, we use the first-principles method, which not only accurately simulates the spectra while distinguishing the ratio of C, H, and Cl elements, but also obtains the gold impurity rate corresponding to various densities. The innovations of this work lie in: (1) We employ the newly proposed ITCF method, using this diagnostic approach to construct a complete set of XRTS diagnostic procedures, decoupling the diagnosis of temperature and density, which greatly reduces the difficulty of diagnosis; (2) In the past, the objects of XRTS diagnosis were basically considered pure, containing only one or two main elements in the target pellet. The diagnosis object of this paper not only contains C, H, and Cl three elements but also confirms the presence of gold impurity. 

Finally, we need to point out: (1) The ITCF method has high requirements for the measurement of XRTS and SIF spectra; only obtaining sufficiently accurate results with minimal noise can ensure the credibility of this method. Given that the time span corresponding to the SIF we utilize is somewhat longer than the duration of the scattering process, it does not represent the true SIF corresponding to XRTS, and this discrepancy may potentially introduce errors; (2) The mixing process of Au exhibits a temporal and spatial distribution. This characteristic could not be accurately diagnosed in the preliminary experiments. Therefore, it is necessary to further design experiments to enhance the temporal and spatial resolution efficiency; (3) Our analysis provides a more accurate method for the diagnosis of the plasma properties at the cone exit. Subsequently, by altering the experimental conditions, we can explore better plasma compression effects and a lower proportion of Au impurity.

\begin{acknowledgments}
This work was supported by the Strategic Priority Research Program of Chinese Academy of Sciences (Grant No. XDA250010100 and XDA250050500), National Natural Science Foundation of China (Grant No. 12075204, No. 12305271 and No. 12247137) and Shanghai Municipal Science and Technology Key Project (Grant No. 22JC1401500). D. Wu thanks the sponsorship from Yangyang Development Fund. J. H. L. thanks the sponsorship from the Shanghai Postdoctoral Excellence Program (No. 2022324). N. Y. S. thanks Yangyang Lin for the useful discussions during the early stages of this work.
\end{acknowledgments}

\appendix

\section{Formula of First-Principles Method and ITCF}

In Eq. (\ref{3}), $S_{ii}(\mathbf{q})$ can be calculated form the Fourier transformation of the ionic pair distribution $g(\mathbf{r})$ as \cite{PhysRevB.102.195127}:
\begin{align}
\begin{split}
   S_{ii}(\mathbf{q})=1+n_i\int[g(\mathbf{r})-1]\exp(i\mathbf{q}\cdot\mathbf{r})d\mathbf{r}
\label{A1}
\end{split}
\end{align}
where $g(\mathbf{r})$ can be  extracted from ionic configurations obtained in the molecular dynamics simulation and $n_i$ is the the average number density of ions. $N(\mathbf{q})$ is determined through the ionic average as \cite{PhysRevB.102.195127}:
\begin{align}
\begin{split}
   N(\mathbf{q})=\langle\frac{\overline{\rho(\mathbf{q})}}{\rho_i(\mathbf{q})}\rangle
\label{A2}
\end{split}
\end{align}
where $\overline{\rho(\mathbf{q})}$ and $\rho_i(\mathbf{q})$ are spatial Fourier transformations of the electronic and ionic densities separately.

In Eq. (\ref{5}), the $\chi_{ee}(\mathbf{q},\omega)$ is the Fourier transformation of the density response function $\chi_{ee}(\mathbf{r},\mathbf{r'})$ in real space, which is the solution of a Dyson-like equation \cite{PhysRevLett.120.205002}:
\begin{align}
\begin{split}
   \chi_{ee}(\mathbf{r},\mathbf{r'},\omega)=\chi_{ee}^0(\mathbf{r},\mathbf{r'},\omega)+\int\mathbf{{dr}_1 {dr}_2}\chi^0_{ee}(\mathbf{r},\mathbf{r_1},\omega) \\\times K(\mathbf{r_1},\mathbf{r_2},\omega)\chi_{ee}(\mathbf{r_2},\mathbf{r'},\omega)
\label{A3}
\end{split}
\end{align}
with $K(\mathbf{r_1},\mathbf{r_2},\omega)=1/|\mathbf{r_1}-\mathbf{r_2}|+f_{xc}^{TD}(\mathbf{r_1},\mathbf{r_2},\omega)$, $f_{xc}^{TD}$ being the time-dependent exchange-correlation kernel. The bare density response function$\chi_{ee}^0$ can be written as \cite{PhysRevLett.120.205002} :
\begin{align}
\begin{split}
   \chi_{ee}^0(\mathbf{r},\mathbf{r'},\omega)=\sum_{j\neq k}(f_j-f_k)\dfrac{\phi_k(\mathbf{r})\phi_j^*(\mathbf{r})\phi_j(\mathbf{r'})\phi_k^*(\mathbf{r'})}{\omega-(\epsilon_j-\epsilon_k)+i\eta}
\label{A4}
\end{split}
\end{align}
where $f_i$ is the Fermi-Dirac occupation of the $i$th level, $\phi_i$ is the corresponding wave function, and $\eta$ is the Lorentzian broadening factor.

Proposed by Dornheim et al. \cite{Dornheim2022}, the ITCF method is a model-free diagnosis for temperature in XRTS. This method not only requires minor calculation, but also avoids the deconvolution of dynamic structure factor $S(\mathbf{q},E)$ with the source-and-instrument function (SIF) $R(E)$, as the spectrum intensity writes $I(\mathbf{q},E)=S(\mathbf{q},E)\otimes R(E)$. The ITCF $F(\mathbf{q},\tau)$ uses the two-sided  Laplace transform of the dynamic structure factor \cite{Dornheim2022}:
\begin{align}
\begin{split}
   F(\mathbf{q},\tau)=\mathcal{L}[S(\mathbf{q},E)]=\int_{-\infty}^{\infty}dEe^{-\tau E}S(\mathbf{q},E)
\label{A5}
\end{split}
\end{align}
where imaginary time $t=-i\hbar\tau\in-i\hbar[0,\beta]$ and $\beta=1/k_B T$ is the inverse temperature. Since the ITCF $F(\mathbf{q},\tau)$ is symmetric around $\tau=\beta/2$, knowledge of $S(\mathbf{q},E)$ directly leads to the exact temperature of the system by calculating the integral and locating the minimum of $F(\mathbf{q},\tau)$ at $\tau=\beta/2$. While in the experiment $S(\mathbf{q},E)$ is always convoluted with SIF, we expand $F(\mathbf{q},\tau)$ as \cite{Dornheim2022,10.1063/5.0222072}:
\begin{align}
\begin{split}
   F(\mathbf{q},\tau)=\frac{\mathcal{L}[S(\mathbf{q},E)\otimes R(E)]}{\mathcal{L}[R(E)]}=\frac{\mathcal{L}[I(\mathbf{q},E)]}{\mathcal{L}[R(E)]}
\label{A6}
\end{split}
\end{align}

\begin{figure}[htbp]
\centering
\includegraphics[scale=0.49]{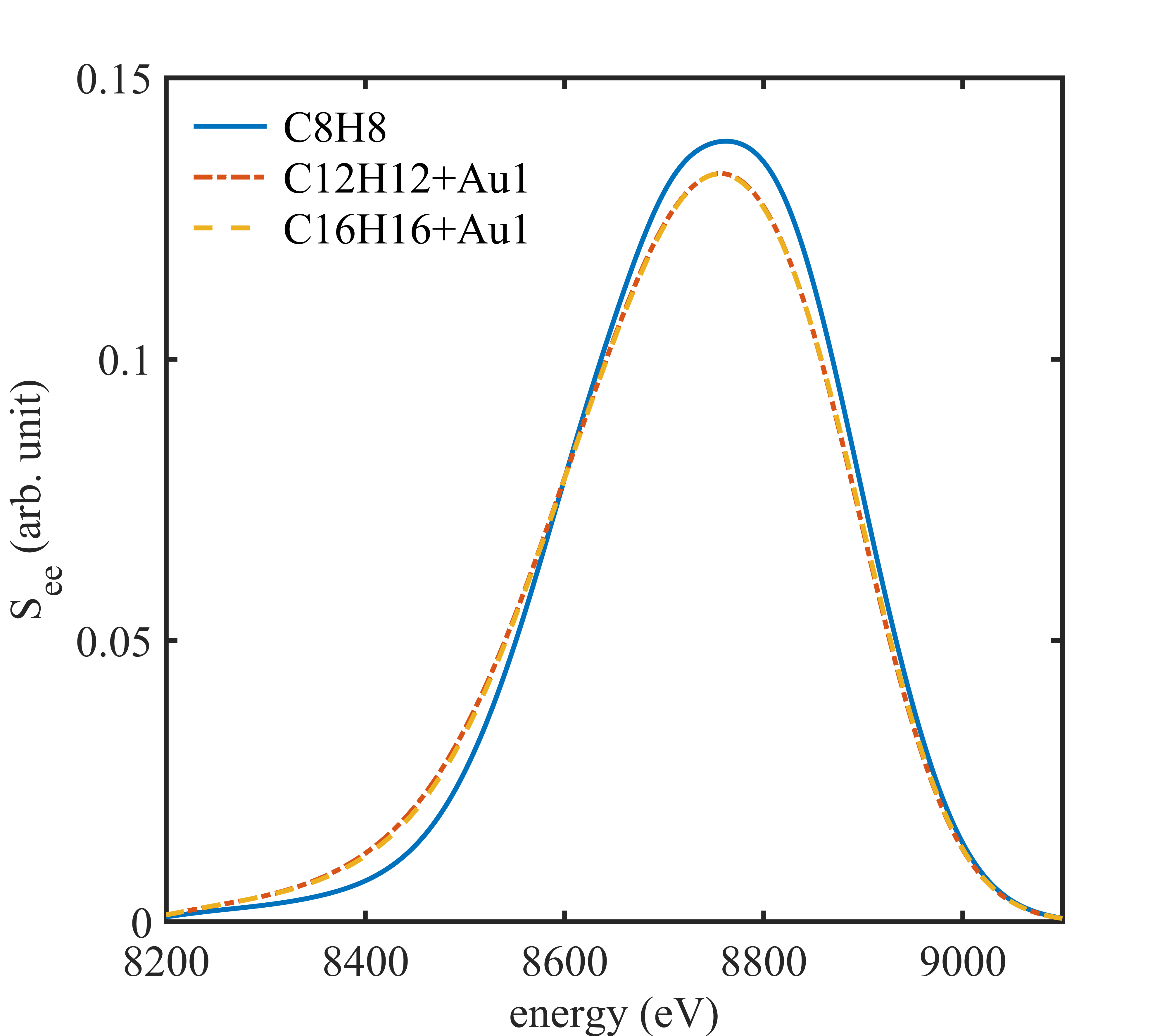}
\caption{The impact of Au on the simulation of the inelastic scattering: three simulations share the same temperature 25 $\mathrm{eV}$ and electron density $4\times10^{24}$ $\mathrm{cm^{-3}}$, while the difference comes from the ratio of Au.}
\label{ine}
\end{figure}

\section{Au Impurity in Free-to-Free Transition}

Fig. \ref{ine} shows the impact of Au on the simulation of the inelastic scattering from free electrons. The temperature and electron density are set to the same in the simulation, i.e. $25$ $\mathrm{eV}$ and $4\times10^{24}$ $\mathrm{cm^{-3}}$. A comparison of the red and yellow lines reveals that the gold impurity ratio has negligible influence on the shape of the inelastic peak. The comparison with the blue line demonstrates that the gold impurity slightly suppresses the peak intensity while inducing a minor elevation of the spectral baseline within the 8400–8500 eV range. These discrepancies may arise from statistical fluctuations near the peak region due to the limited number of atoms, which could introduce uncertainty in the estimated gold impurity ratio. Nevertheless, given the exceptionally low gold doping level and the computational constraints associated with modeling hundreds of atoms, we propose that gold can be reasonably disregarded in free-to-free inelastic scattering calculations.

\section{Calculations in Bound-to-Bound Transition}

\begin{figure}[H]
\centering
\includegraphics[scale=0.49]{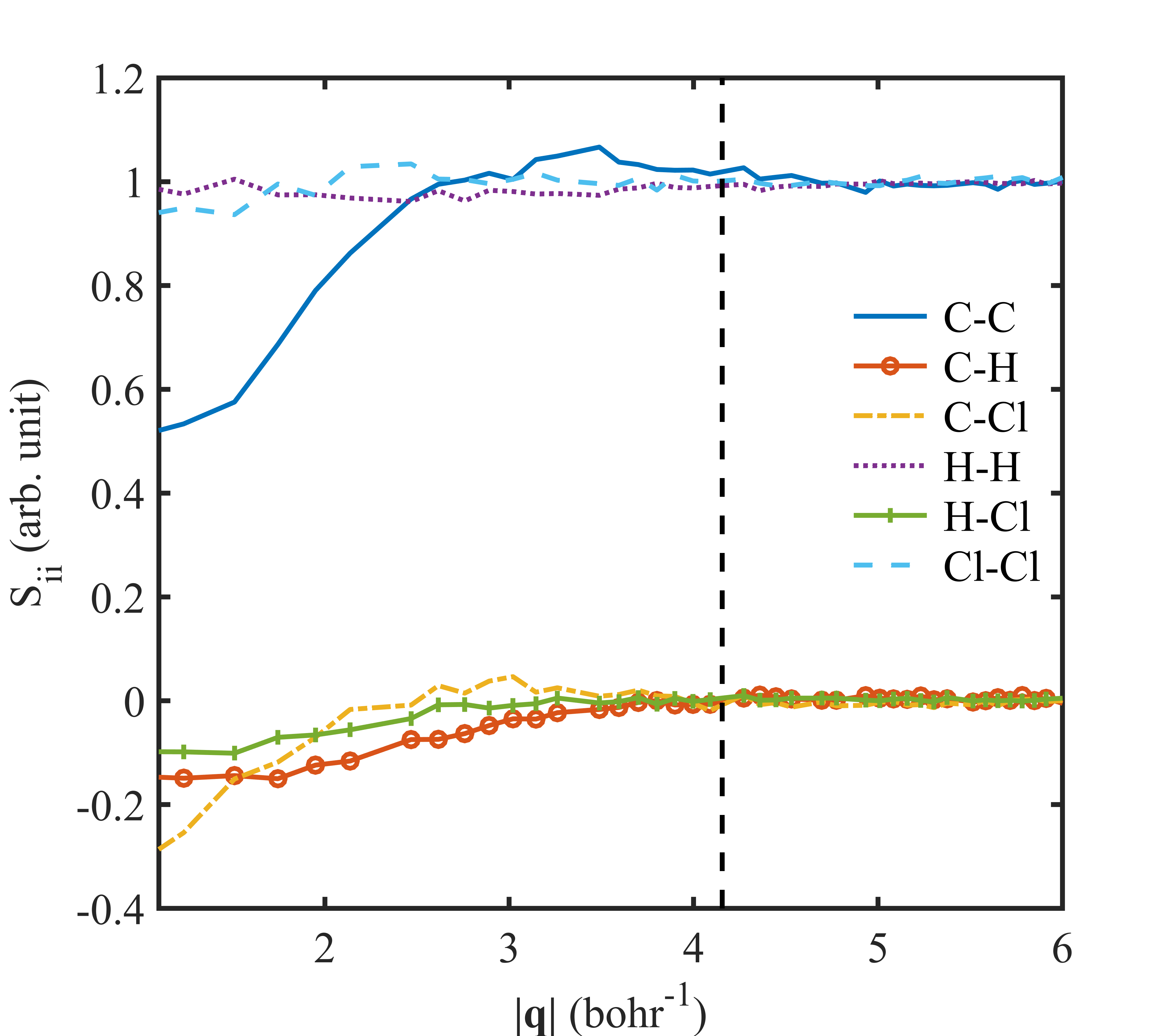}
\caption{ $S_{\mu\nu}(\mathbf{q})$ under 25 eV and 8 g/cc: six solid lines display $S_{\mu\nu}(\mathbf{q})$ of different component $\mu$ and $\nu$, while black dotted line shows the transferred momentum $\lvert \mathbf{q} \rvert=4.16$ $\mathrm{bohr}^{-1}$.}
\label{Sii}
\end{figure}

In our treatment of bound-bound elastic scattering, we employ a tightly-bound electron model to derive electronic form factors through orbital wave functions under zero temperature. This approximation was adopted because both electronic form factors and static ionic structure factors require averaged ionic configurations: a procedure rendered impractical by the minimal gold proportion in molecular dynamic simulations. While this approximation neglects temperature effects on eigen-wavefunctions and higher excited states, it remains well-justified given the dominant contributions from inner-shell electrons.
\begin{figure}[H]
\centering
\includegraphics[scale=0.49]{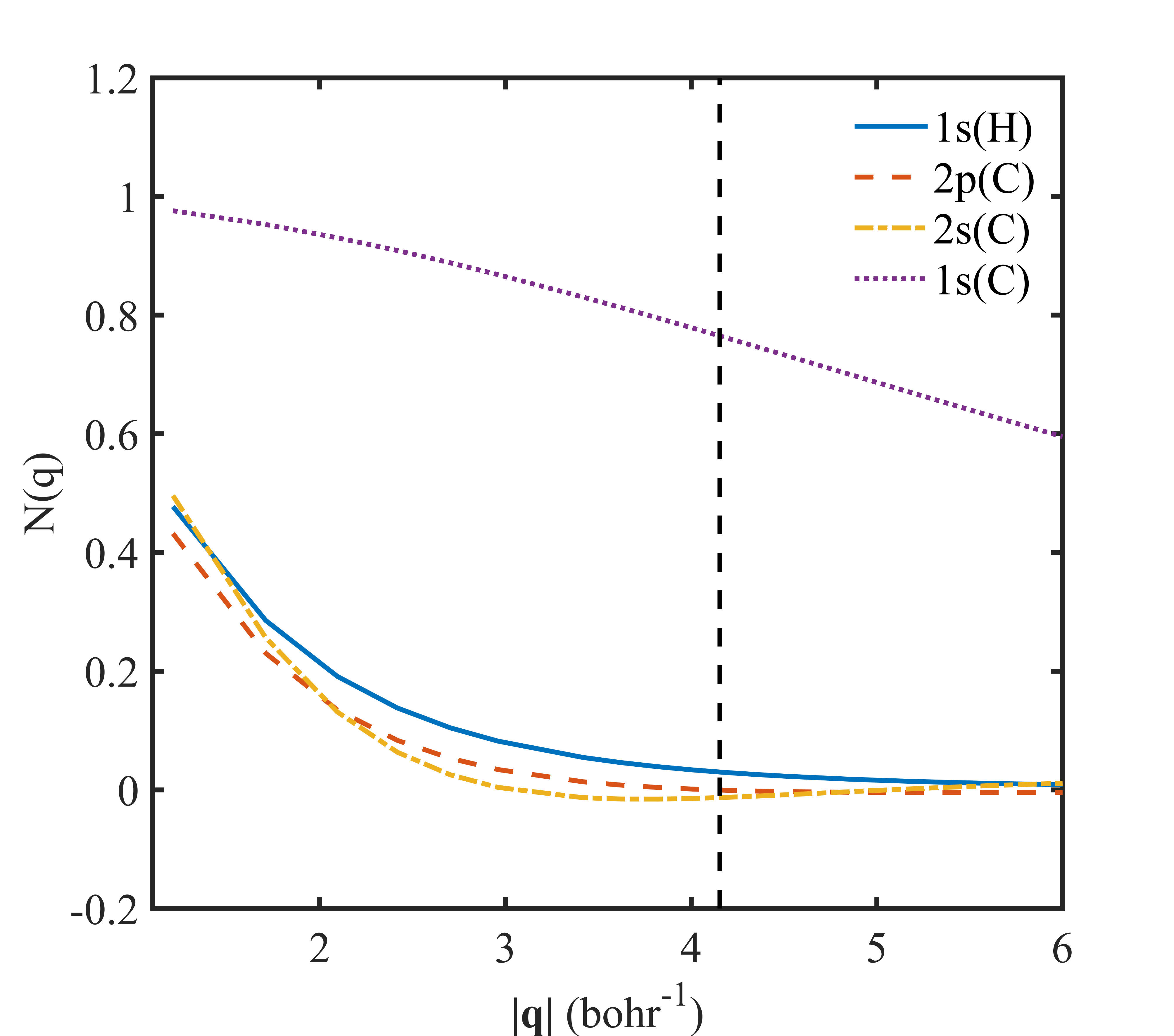}
\caption{$N(\mathbf{q})$ for H and C: Black dotted line shows the transferred momentum $\lvert \mathbf{q} \rvert =4.16$ $\mathrm{bohr}^{-1}$. Solid line stands for H and dotted lines stand for C.}
\label{H-C}
\end{figure}
In calculation of the elastic peak, $S_{\mu\nu}(\mathbf{q})$ and $N(\mathbf{q})$ are shown in the following figures. Black dotted line shows the transferred momentum $\lvert \mathbf{q} \rvert =4.16$ $\mathrm{bohr}^{-1}$. Fig. \ref{Sii} provides a basis for the approximation in gold impurity, as the $\mu$-$\mu$ components approach 1, and the $\mu$-$\nu$ components approach 0. In addition, the impact produced by $N(\mathbf{q})$ is much greater than that produced by $S_{\mu\nu}(\mathbf{q})$ in our simulation due to the inner bound electrons of Cl and Au.

\begin{figure}[H]
\centering
\includegraphics[scale=0.49]{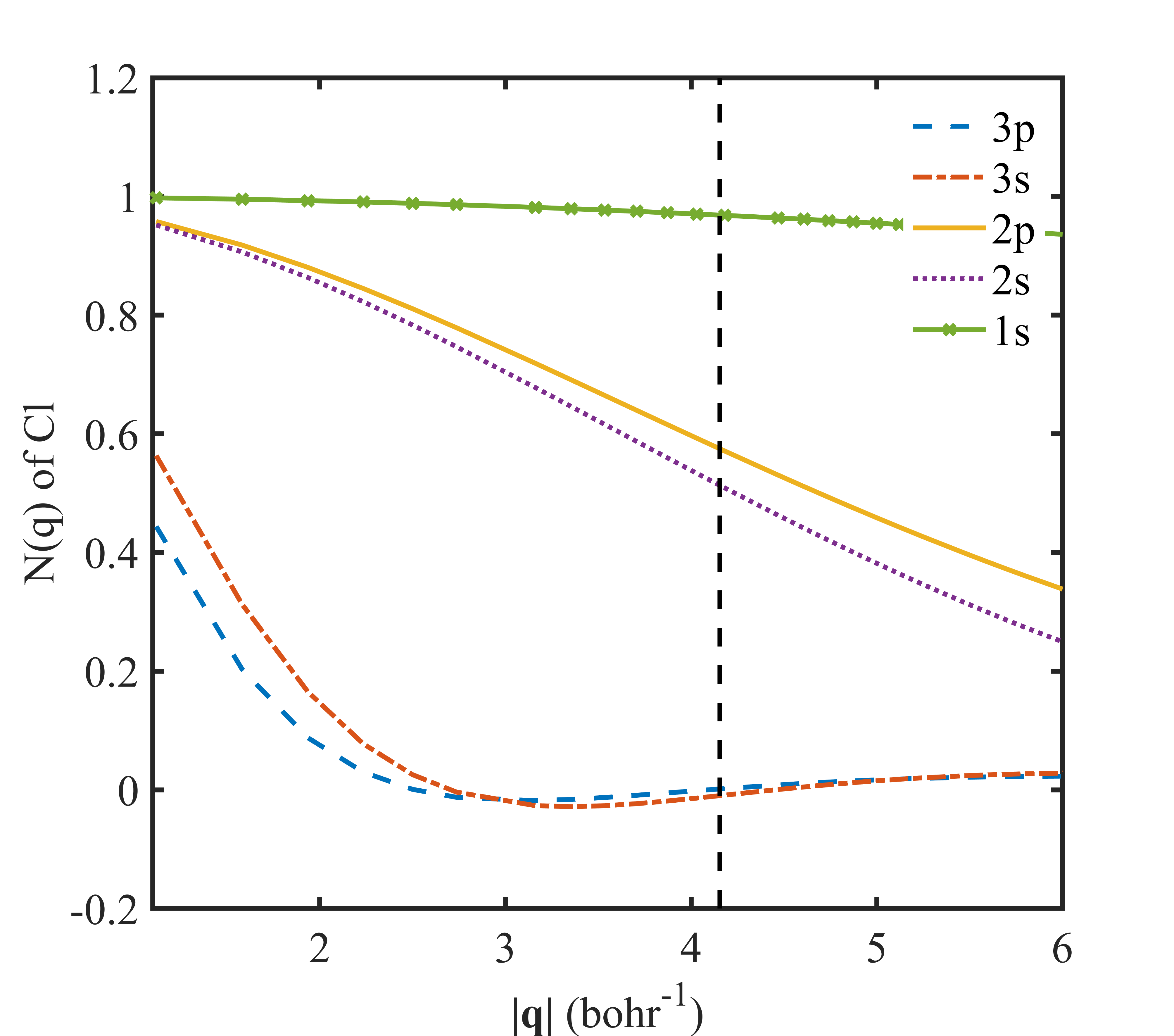}
\caption{$N(\mathbf{q})$ for Cl: Black dotted line shows the transferred momentum $\lvert \mathbf{q} \rvert =4.16$ $\mathrm{bohr}^{-1}$.}
\label{Cl}
\end{figure}

\begin{figure}[H]
\centering
\includegraphics[scale=0.49]{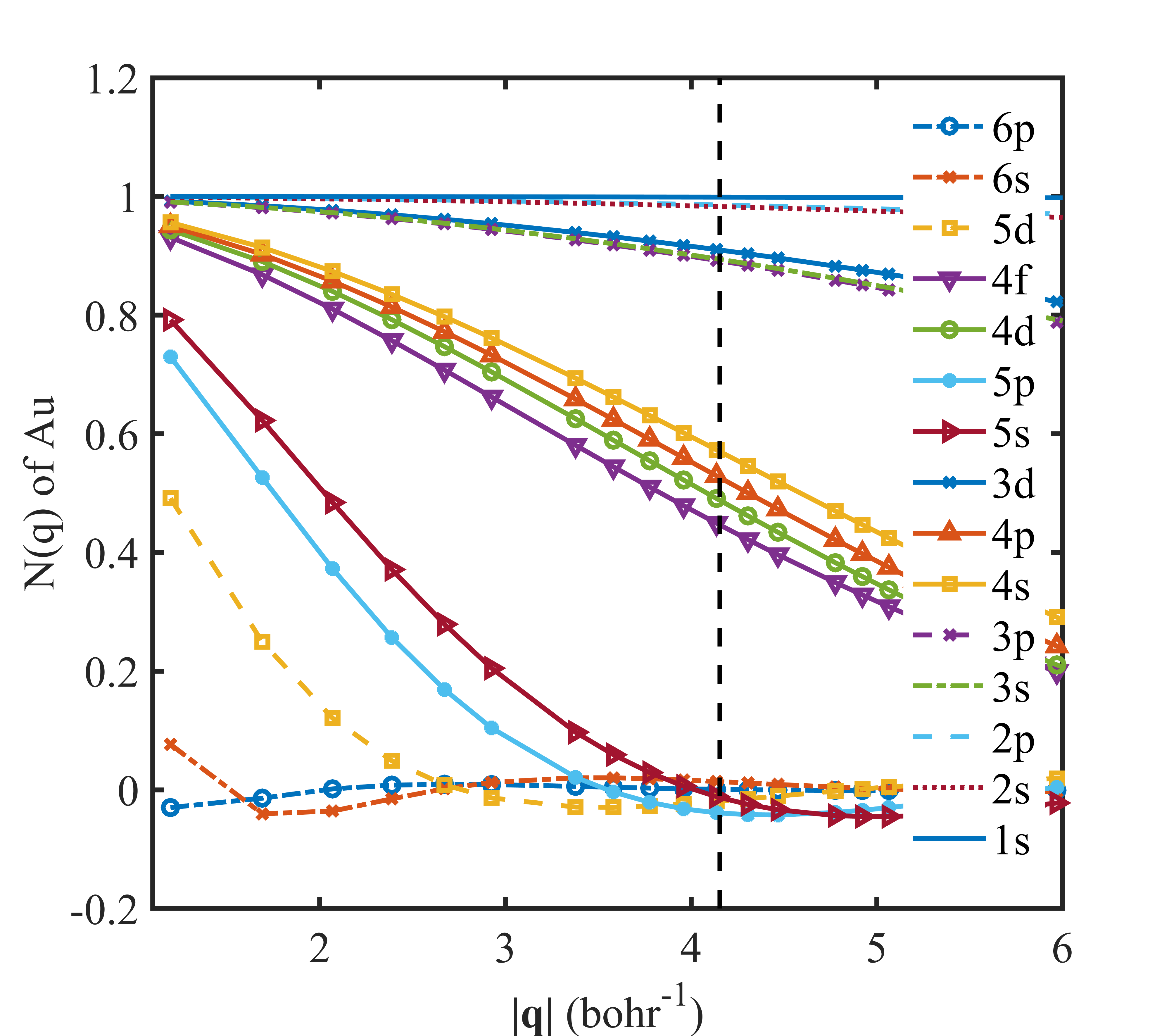}
\caption{$N(\mathbf{q})$ for Au: Black dotted line shows the transferred momentum $\lvert \mathbf{q} \rvert =4.16$ $\mathrm{bohr}^{-1}$.}
\label{Au}
\end{figure}

Fig. \ref{H-C}, \ref{Cl} and \ref{Au} illustrate the $\mathrm{N(\mathbf{q})}$ values of various elements on different orbits. Fig. \ref{H-C} illustrates $\mathrm{N(\mathbf{q})}$ of H and C, Fig. \ref{Cl} illustrates $\mathrm{N(\mathbf{q})}$ of Cl and Fig. \ref{Au} illustrates $\mathrm{N(\mathbf{q})}$ of Au. By calculating the occupation numbers of each orbit for element $\mu$ via Quantum Espresso, the corresponding $N_\mu(\mathbf{q})$ for element $\mu$ can be obtained.

\nocite{*}
\bibliography{aipsamp}

\begin{thebibliography}{25}%
\makeatletter
\providecommand \@ifxundefined [1]{%
 \@ifx{#1\undefined}
}%
\providecommand \@ifnum [1]{%
 \ifnum #1\expandafter \@firstoftwo
 \else \expandafter \@secondoftwo
 \fi
}%
\providecommand \@ifx [1]{%
 \ifx #1\expandafter \@firstoftwo
 \else \expandafter \@secondoftwo
 \fi
}%
\providecommand \natexlab [1]{#1}%
\providecommand \enquote  [1]{``#1''}%
\providecommand \bibnamefont  [1]{#1}%
\providecommand \bibfnamefont [1]{#1}%
\providecommand \citenamefont [1]{#1}%
\providecommand \href@noop [0]{\@secondoftwo}%
\providecommand \href [0]{\begingroup \@sanitize@url \@href}%
\providecommand \@href[1]{\@@startlink{#1}\@@href}%
\providecommand \@@href[1]{\endgroup#1\@@endlink}%
\providecommand \@sanitize@url [0]{\catcode `\\12\catcode `\$12\catcode
  `\&12\catcode `\#12\catcode `\^12\catcode `\_12\catcode `\%12\relax}%
\providecommand \@@startlink[1]{}%
\providecommand \@@endlink[0]{}%
\providecommand \url  [0]{\begingroup\@sanitize@url \@url }%
\providecommand \@url [1]{\endgroup\@href {#1}{\urlprefix }}%
\providecommand \urlprefix  [0]{URL }%
\providecommand \Eprint [0]{\href }%
\providecommand \doibase [0]{http://dx.doi.org/}%
\providecommand \selectlanguage [0]{\@gobble}%
\providecommand \bibinfo  [0]{\@secondoftwo}%
\providecommand \bibfield  [0]{\@secondoftwo}%
\providecommand \translation [1]{[#1]}%
\providecommand \BibitemOpen [0]{}%
\providecommand \bibitemStop [0]{}%
\providecommand \bibitemNoStop [0]{.\EOS\space}%
\providecommand \EOS [0]{\spacefactor3000\relax}%
\providecommand \BibitemShut  [1]{\csname bibitem#1\endcsname}%
\let\auto@bib@innerbib\@empty
\bibitem [{\citenamefont {Zhang}\ \emph {et~al.}(2020)\citenamefont {Zhang},
  \citenamefont {Wang}, \citenamefont {Yang}, \citenamefont {Wu}, \citenamefont
  {Ma}, \citenamefont {Jiao}, \citenamefont {Zhang}, \citenamefont {Wu},
  \citenamefont {Yuan}, \citenamefont {Li},\ and\ \citenamefont
  {Zhu}}]{zhang2020double}%
  \BibitemOpen
  \bibfield  {author} {\bibinfo {author} {\bibfnamefont {J.}~\bibnamefont
  {Zhang}}, \bibinfo {author} {\bibfnamefont {W.~M.}\ \bibnamefont {Wang}},
  \bibinfo {author} {\bibfnamefont {X.~H.}\ \bibnamefont {Yang}}, \bibinfo
  {author} {\bibfnamefont {D.}~\bibnamefont {Wu}}, \bibinfo {author}
  {\bibfnamefont {Y.~Y.}\ \bibnamefont {Ma}}, \bibinfo {author} {\bibfnamefont
  {J.~L.}\ \bibnamefont {Jiao}}, \bibinfo {author} {\bibfnamefont
  {Z.}~\bibnamefont {Zhang}}, \bibinfo {author} {\bibfnamefont {F.~Y.}\
  \bibnamefont {Wu}}, \bibinfo {author} {\bibfnamefont {X.~H.}\ \bibnamefont
  {Yuan}}, \bibinfo {author} {\bibfnamefont {Y.~T.}\ \bibnamefont {Li}}, \ and\
  \bibinfo {author} {\bibfnamefont {J.~Q.}\ \bibnamefont {Zhu}},\ }\bibfield
  {title} {\enquote {\bibinfo {title} {Double-cone ignition scheme for inertial
  confinement fusion},}\ }\href@noop {} {\bibfield  {journal} {\bibinfo
  {journal} {Philosophical Transactions of the Royal Society A}\ }\textbf
  {\bibinfo {volume} {378}},\ \bibinfo {pages} {015} (\bibinfo {year}
  {2020})}\BibitemShut {NoStop}%
\bibitem [{\citenamefont {Molitoris}, \citenamefont {Lee},\ and\ \citenamefont
  {Kalantar}(2004)}]{unknown}%
  \BibitemOpen
  \bibfield  {author} {\bibinfo {author} {\bibfnamefont {J.}~\bibnamefont
  {Molitoris}}, \bibinfo {author} {\bibfnamefont {R.}~\bibnamefont {Lee}}, \
  and\ \bibinfo {author} {\bibfnamefont {D.}~\bibnamefont {Kalantar}},\ }\href
  {\doibase 10.13140/RG.2.1.4037.2722} {\enquote {\bibinfo {title} {Warm dense
  matter: An overview},}\ } (\bibinfo {year} {2004})\BibitemShut {NoStop}%
\bibitem [{\citenamefont {Lahmann}\ \emph {et~al.}(2023)\citenamefont
  {Lahmann}, \citenamefont {Saunders}, \citenamefont {Döppner}, \citenamefont
  {Frenje}, \citenamefont {Glenzer}, \citenamefont {Gatu-Johnson},
  \citenamefont {Sutcliffe}, \citenamefont {Zylstra},\ and\ \citenamefont
  {Petrasso}}]{Lahmann_2023}%
  \BibitemOpen
  \bibfield  {author} {\bibinfo {author} {\bibfnamefont {B.}~\bibnamefont
  {Lahmann}}, \bibinfo {author} {\bibfnamefont {A.~M.}\ \bibnamefont
  {Saunders}}, \bibinfo {author} {\bibfnamefont {T.}~\bibnamefont {Döppner}},
  \bibinfo {author} {\bibfnamefont {J.~A.}\ \bibnamefont {Frenje}}, \bibinfo
  {author} {\bibfnamefont {S.~H.}\ \bibnamefont {Glenzer}}, \bibinfo {author}
  {\bibfnamefont {M.}~\bibnamefont {Gatu-Johnson}}, \bibinfo {author}
  {\bibfnamefont {G.}~\bibnamefont {Sutcliffe}}, \bibinfo {author}
  {\bibfnamefont {A.~B.}\ \bibnamefont {Zylstra}}, \ and\ \bibinfo {author}
  {\bibfnamefont {R.~D.}\ \bibnamefont {Petrasso}},\ }\bibfield  {title}
  {\enquote {\bibinfo {title} {Measuring stopping power in warm dense matter
  plasmas at omega},}\ }\href {\doibase 10.1088/1361-6587/ace4f2} {\bibfield
  {journal} {\bibinfo  {journal} {Plasma Physics and Controlled Fusion}\
  }\textbf {\bibinfo {volume} {65}},\ \bibinfo {pages} {095002} (\bibinfo
  {year} {2023})}\BibitemShut {NoStop}%
\bibitem [{\citenamefont {Graziani}\ \emph {et~al.}(2014)\citenamefont
  {Graziani}, \citenamefont {Desjarlais}, \citenamefont {Redmer},\ and\
  \citenamefont {Trickey}}]{graziani2014frontiers}%
  \BibitemOpen
  \bibfield  {author} {\bibinfo {author} {\bibfnamefont {F.}~\bibnamefont
  {Graziani}}, \bibinfo {author} {\bibfnamefont {M.~P.}\ \bibnamefont
  {Desjarlais}}, \bibinfo {author} {\bibfnamefont {R.}~\bibnamefont {Redmer}},
  \ and\ \bibinfo {author} {\bibfnamefont {S.~B.}\ \bibnamefont {Trickey}},\
  }\href {\doibase 10.1007/978-3-319-04912-0} {\emph {\bibinfo {title}
  {Frontiers and Challenges in Warm Dense Matter}}},\ Lecture Notes in
  Computational Science and Engineering\ (\bibinfo  {publisher} {Springer
  Cham},\ \bibinfo {year} {2014})\ pp.\ \bibinfo {pages} {X, 282}\BibitemShut
  {NoStop}%
\bibitem [{\citenamefont {Saunders}\ \emph {et~al.}(2018)\citenamefont
  {Saunders}, \citenamefont {Chapman}, \citenamefont {Kritcher}, \citenamefont
  {Schoff}, \citenamefont {Shuldberg}, \citenamefont {Landen}, \citenamefont
  {Glenzer}, \citenamefont {Falcone}, \citenamefont {Gericke},\ and\
  \citenamefont {Döppner}}]{SAUNDERS201886}%
  \BibitemOpen
  \bibfield  {author} {\bibinfo {author} {\bibfnamefont {A.}~\bibnamefont
  {Saunders}}, \bibinfo {author} {\bibfnamefont {D.}~\bibnamefont {Chapman}},
  \bibinfo {author} {\bibfnamefont {A.}~\bibnamefont {Kritcher}}, \bibinfo
  {author} {\bibfnamefont {M.}~\bibnamefont {Schoff}}, \bibinfo {author}
  {\bibfnamefont {C.}~\bibnamefont {Shuldberg}}, \bibinfo {author}
  {\bibfnamefont {O.}~\bibnamefont {Landen}}, \bibinfo {author} {\bibfnamefont
  {S.}~\bibnamefont {Glenzer}}, \bibinfo {author} {\bibfnamefont
  {R.}~\bibnamefont {Falcone}}, \bibinfo {author} {\bibfnamefont
  {D.}~\bibnamefont {Gericke}}, \ and\ \bibinfo {author} {\bibfnamefont
  {T.}~\bibnamefont {Döppner}},\ }\bibfield  {title} {\enquote {\bibinfo
  {title} {Influence of argon impurities on the elastic scattering of x-rays
  from imploding beryllium capsules},}\ }\href {\doibase
  https://doi.org/10.1016/j.hedp.2018.02.003} {\bibfield  {journal} {\bibinfo
  {journal} {High Energy Density Physics}\ }\textbf {\bibinfo {volume} {26}},\
  \bibinfo {pages} {86--92} (\bibinfo {year} {2018})}\BibitemShut {NoStop}%
\bibitem [{\citenamefont {Li}\ \emph {et~al.}(2022)\citenamefont {Li},
  \citenamefont {Zhang}, \citenamefont {Chen} \emph {et~al.}}]{Li2022}%
  \BibitemOpen
  \bibfield  {author} {\bibinfo {author} {\bibfnamefont {M.}~\bibnamefont
  {Li}}, \bibinfo {author} {\bibfnamefont {H.}~\bibnamefont {Zhang}}, \bibinfo
  {author} {\bibfnamefont {S.}~\bibnamefont {Chen}},  \emph {et~al.},\
  }\bibfield  {title} {\enquote {\bibinfo {title} {Laser driven dynamic
  compression of materials},}\ }\href {\doibase 10.11884/HPLPB202234.210357}
  {\bibfield  {journal} {\bibinfo  {journal} {High Power Laser and Particle
  Beams}\ }\textbf {\bibinfo {volume} {34}},\ \bibinfo {pages} {011001}
  (\bibinfo {year} {2022})}\BibitemShut {NoStop}%
\bibitem [{\citenamefont {Haack}, \citenamefont {Hauck},\ and\ \citenamefont
  {Murillo}(2017)}]{PhysRevE.96.063310}%
  \BibitemOpen
  \bibfield  {author} {\bibinfo {author} {\bibfnamefont {J.~R.}\ \bibnamefont
  {Haack}}, \bibinfo {author} {\bibfnamefont {C.~D.}\ \bibnamefont {Hauck}}, \
  and\ \bibinfo {author} {\bibfnamefont {M.~S.}\ \bibnamefont {Murillo}},\
  }\bibfield  {title} {\enquote {\bibinfo {title} {Interfacial mixing in
  high-energy-density matter with a multiphysics kinetic model},}\ }\href
  {\doibase 10.1103/PhysRevE.96.063310} {\bibfield  {journal} {\bibinfo
  {journal} {Phys. Rev. E}\ }\textbf {\bibinfo {volume} {96}},\ \bibinfo
  {pages} {063310} (\bibinfo {year} {2017})}\BibitemShut {NoStop}%
\bibitem [{\citenamefont {Mahieu}\ \emph {et~al.}(2018)\citenamefont {Mahieu},
  \citenamefont {Jourdain}, \citenamefont {Ta~Phuoc}, \citenamefont {Dorchies},
  \citenamefont {Goddet}, \citenamefont {Lifschitz}, \citenamefont {Renaudin},\
  and\ \citenamefont {Lecherbourg}}]{Mahieu2018}%
  \BibitemOpen
  \bibfield  {author} {\bibinfo {author} {\bibfnamefont {B.}~\bibnamefont
  {Mahieu}}, \bibinfo {author} {\bibfnamefont {N.}~\bibnamefont {Jourdain}},
  \bibinfo {author} {\bibfnamefont {K.}~\bibnamefont {Ta~Phuoc}}, \bibinfo
  {author} {\bibfnamefont {F.}~\bibnamefont {Dorchies}}, \bibinfo {author}
  {\bibfnamefont {J.-P.}\ \bibnamefont {Goddet}}, \bibinfo {author}
  {\bibfnamefont {A.}~\bibnamefont {Lifschitz}}, \bibinfo {author}
  {\bibfnamefont {P.}~\bibnamefont {Renaudin}}, \ and\ \bibinfo {author}
  {\bibfnamefont {L.}~\bibnamefont {Lecherbourg}},\ }\bibfield  {title}
  {\enquote {\bibinfo {title} {Probing warm dense matter using femtosecond
  x-ray absorption spectroscopy with a laser-produced betatron source},}\
  }\href {\doibase 10.1038/s41467-018-05791-4} {\bibfield  {journal} {\bibinfo
  {journal} {Nature Communications}\ }\textbf {\bibinfo {volume} {9}},\
  \bibinfo {pages} {3276} (\bibinfo {year} {2018})}\BibitemShut {NoStop}%
\bibitem [{\citenamefont {Stoeckl}\ \emph {et~al.}(2014)\citenamefont
  {Stoeckl}, \citenamefont {Bedzyk}, \citenamefont {Brent}, \citenamefont
  {Epstein}, \citenamefont {Fiksel}, \citenamefont {Guy}, \citenamefont
  {Goncharov}, \citenamefont {Hu}, \citenamefont {Ingraham}, \citenamefont
  {Jacobs-Perkins}, \citenamefont {Jungquist}, \citenamefont {Marshall},
  \citenamefont {Mileham}, \citenamefont {Nilson}, \citenamefont {Sangster},
  \citenamefont {Shoup},\ and\ \citenamefont {Theobald}}]{10.1063/1.4890215}%
  \BibitemOpen
  \bibfield  {author} {\bibinfo {author} {\bibfnamefont {C.}~\bibnamefont
  {Stoeckl}}, \bibinfo {author} {\bibfnamefont {M.}~\bibnamefont {Bedzyk}},
  \bibinfo {author} {\bibfnamefont {G.}~\bibnamefont {Brent}}, \bibinfo
  {author} {\bibfnamefont {R.}~\bibnamefont {Epstein}}, \bibinfo {author}
  {\bibfnamefont {G.}~\bibnamefont {Fiksel}}, \bibinfo {author} {\bibfnamefont
  {D.}~\bibnamefont {Guy}}, \bibinfo {author} {\bibfnamefont {V.~N.}\
  \bibnamefont {Goncharov}}, \bibinfo {author} {\bibfnamefont {S.~X.}\
  \bibnamefont {Hu}}, \bibinfo {author} {\bibfnamefont {S.}~\bibnamefont
  {Ingraham}}, \bibinfo {author} {\bibfnamefont {D.~W.}\ \bibnamefont
  {Jacobs-Perkins}}, \bibinfo {author} {\bibfnamefont {R.~K.}\ \bibnamefont
  {Jungquist}}, \bibinfo {author} {\bibfnamefont {F.~J.}\ \bibnamefont
  {Marshall}}, \bibinfo {author} {\bibfnamefont {C.}~\bibnamefont {Mileham}},
  \bibinfo {author} {\bibfnamefont {P.~M.}\ \bibnamefont {Nilson}}, \bibinfo
  {author} {\bibfnamefont {T.~C.}\ \bibnamefont {Sangster}}, \bibinfo {author}
  {\bibfnamefont {I.}~\bibnamefont {Shoup}, \bibfnamefont {M.~J.}}, \ and\
  \bibinfo {author} {\bibfnamefont {W.}~\bibnamefont {Theobald}},\ }\bibfield
  {title} {\enquote {\bibinfo {title} {{Soft x-ray backlighting of cryogenic
  implosions using a narrowband crystal imaging system (invited)a)}},}\ }\href
  {\doibase 10.1063/1.4890215} {\bibfield  {journal} {\bibinfo  {journal}
  {Review of Scientific Instruments}\ }\textbf {\bibinfo {volume} {85}},\
  \bibinfo {pages} {11E501} (\bibinfo {year} {2014})}\BibitemShut {NoStop}%
\bibitem [{\citenamefont {Sawada}\ \emph {et~al.}(2017)\citenamefont {Sawada},
  \citenamefont {Daykin}, \citenamefont {McLean}, \citenamefont {Chen},
  \citenamefont {Patel}, \citenamefont {Ping},\ and\ \citenamefont
  {Pérez}}]{10.1063/1.4985729}%
  \BibitemOpen
  \bibfield  {author} {\bibinfo {author} {\bibfnamefont {H.}~\bibnamefont
  {Sawada}}, \bibinfo {author} {\bibfnamefont {T.}~\bibnamefont {Daykin}},
  \bibinfo {author} {\bibfnamefont {H.~S.}\ \bibnamefont {McLean}}, \bibinfo
  {author} {\bibfnamefont {H.}~\bibnamefont {Chen}}, \bibinfo {author}
  {\bibfnamefont {P.~K.}\ \bibnamefont {Patel}}, \bibinfo {author}
  {\bibfnamefont {Y.}~\bibnamefont {Ping}}, \ and\ \bibinfo {author}
  {\bibfnamefont {F.}~\bibnamefont {Pérez}},\ }\bibfield  {title} {\enquote
  {\bibinfo {title} {{Two-color monochromatic x-ray imaging with a single
  short-pulse laser}},}\ }\href {\doibase 10.1063/1.4985729} {\bibfield
  {journal} {\bibinfo  {journal} {Review of Scientific Instruments}\ }\textbf
  {\bibinfo {volume} {88}},\ \bibinfo {pages} {063502} (\bibinfo {year}
  {2017})}\BibitemShut {NoStop}%
\bibitem [{\citenamefont {Glenzer}\ and\ \citenamefont
  {Redmer}(2009)}]{RevModPhys.81.1625}%
  \BibitemOpen
  \bibfield  {author} {\bibinfo {author} {\bibfnamefont {S.~H.}\ \bibnamefont
  {Glenzer}}\ and\ \bibinfo {author} {\bibfnamefont {R.}~\bibnamefont
  {Redmer}},\ }\bibfield  {title} {\enquote {\bibinfo {title} {X-ray thomson
  scattering in high energy density plasmas},}\ }\href {\doibase
  10.1103/RevModPhys.81.1625} {\bibfield  {journal} {\bibinfo  {journal} {Rev.
  Mod. Phys.}\ }\textbf {\bibinfo {volume} {81}},\ \bibinfo {pages}
  {1625--1663} (\bibinfo {year} {2009})}\BibitemShut {NoStop}%
\bibitem [{\citenamefont {Sperling}\ \emph {et~al.}(2015)\citenamefont
  {Sperling}, \citenamefont {Gamboa}, \citenamefont {Lee}, \citenamefont
  {Chung}, \citenamefont {Galtier}, \citenamefont {Omarbakiyeva}, \citenamefont
  {Reinholz}, \citenamefont {R\"opke}, \citenamefont {Zastrau}, \citenamefont
  {Hastings}, \citenamefont {Fletcher},\ and\ \citenamefont
  {Glenzer}}]{PhysRevLett.115.115001}%
  \BibitemOpen
  \bibfield  {author} {\bibinfo {author} {\bibfnamefont {P.}~\bibnamefont
  {Sperling}}, \bibinfo {author} {\bibfnamefont {E.~J.}\ \bibnamefont
  {Gamboa}}, \bibinfo {author} {\bibfnamefont {H.~J.}\ \bibnamefont {Lee}},
  \bibinfo {author} {\bibfnamefont {H.~K.}\ \bibnamefont {Chung}}, \bibinfo
  {author} {\bibfnamefont {E.}~\bibnamefont {Galtier}}, \bibinfo {author}
  {\bibfnamefont {Y.}~\bibnamefont {Omarbakiyeva}}, \bibinfo {author}
  {\bibfnamefont {H.}~\bibnamefont {Reinholz}}, \bibinfo {author}
  {\bibfnamefont {G.}~\bibnamefont {R\"opke}}, \bibinfo {author} {\bibfnamefont
  {U.}~\bibnamefont {Zastrau}}, \bibinfo {author} {\bibfnamefont
  {J.}~\bibnamefont {Hastings}}, \bibinfo {author} {\bibfnamefont {L.~B.}\
  \bibnamefont {Fletcher}}, \ and\ \bibinfo {author} {\bibfnamefont {S.~H.}\
  \bibnamefont {Glenzer}},\ }\bibfield  {title} {\enquote {\bibinfo {title}
  {Free-electron x-ray laser measurements of collisional-damped plasmons in
  isochorically heated warm dense matter},}\ }\href {\doibase
  10.1103/PhysRevLett.115.115001} {\bibfield  {journal} {\bibinfo  {journal}
  {Phys. Rev. Lett.}\ }\textbf {\bibinfo {volume} {115}},\ \bibinfo {pages}
  {115001} (\bibinfo {year} {2015})}\BibitemShut {NoStop}%
\bibitem [{\citenamefont {Kritcher}\ \emph {et~al.}(2011)\citenamefont
  {Kritcher}, \citenamefont {D\"oppner}, \citenamefont {Fortmann},
  \citenamefont {Ma}, \citenamefont {Landen}, \citenamefont {Wallace},\ and\
  \citenamefont {Glenzer}}]{PhysRevLett.107.015002}%
  \BibitemOpen
  \bibfield  {author} {\bibinfo {author} {\bibfnamefont {A.~L.}\ \bibnamefont
  {Kritcher}}, \bibinfo {author} {\bibfnamefont {T.}~\bibnamefont {D\"oppner}},
  \bibinfo {author} {\bibfnamefont {C.}~\bibnamefont {Fortmann}}, \bibinfo
  {author} {\bibfnamefont {T.}~\bibnamefont {Ma}}, \bibinfo {author}
  {\bibfnamefont {O.~L.}\ \bibnamefont {Landen}}, \bibinfo {author}
  {\bibfnamefont {R.}~\bibnamefont {Wallace}}, \ and\ \bibinfo {author}
  {\bibfnamefont {S.~H.}\ \bibnamefont {Glenzer}},\ }\bibfield  {title}
  {\enquote {\bibinfo {title} {In-flight measurements of capsule shell adiabats
  in laser-driven implosions},}\ }\href {\doibase
  10.1103/PhysRevLett.107.015002} {\bibfield  {journal} {\bibinfo  {journal}
  {Phys. Rev. Lett.}\ }\textbf {\bibinfo {volume} {107}},\ \bibinfo {pages}
  {015002} (\bibinfo {year} {2011})}\BibitemShut {NoStop}%
\bibitem [{\citenamefont {Fletcher}\ \emph {et~al.}(2014)\citenamefont
  {Fletcher}, \citenamefont {Kritcher}, \citenamefont {Pak}, \citenamefont
  {Ma}, \citenamefont {D\"oppner}, \citenamefont {Fortmann}, \citenamefont
  {Divol}, \citenamefont {Jones}, \citenamefont {Landen}, \citenamefont
  {Scott}, \citenamefont {Vorberger}, \citenamefont {Chapman}, \citenamefont
  {Gericke}, \citenamefont {Mattern}, \citenamefont {Seidler}, \citenamefont
  {Gregori}, \citenamefont {Falcone},\ and\ \citenamefont
  {Glenzer}}]{PhysRevLett.112.145004}%
  \BibitemOpen
  \bibfield  {author} {\bibinfo {author} {\bibfnamefont {L.~B.}\ \bibnamefont
  {Fletcher}}, \bibinfo {author} {\bibfnamefont {A.~L.}\ \bibnamefont
  {Kritcher}}, \bibinfo {author} {\bibfnamefont {A.}~\bibnamefont {Pak}},
  \bibinfo {author} {\bibfnamefont {T.}~\bibnamefont {Ma}}, \bibinfo {author}
  {\bibfnamefont {T.}~\bibnamefont {D\"oppner}}, \bibinfo {author}
  {\bibfnamefont {C.}~\bibnamefont {Fortmann}}, \bibinfo {author}
  {\bibfnamefont {L.}~\bibnamefont {Divol}}, \bibinfo {author} {\bibfnamefont
  {O.~S.}\ \bibnamefont {Jones}}, \bibinfo {author} {\bibfnamefont {O.~L.}\
  \bibnamefont {Landen}}, \bibinfo {author} {\bibfnamefont {H.~A.}\
  \bibnamefont {Scott}}, \bibinfo {author} {\bibfnamefont {J.}~\bibnamefont
  {Vorberger}}, \bibinfo {author} {\bibfnamefont {D.~A.}\ \bibnamefont
  {Chapman}}, \bibinfo {author} {\bibfnamefont {D.~O.}\ \bibnamefont
  {Gericke}}, \bibinfo {author} {\bibfnamefont {B.~A.}\ \bibnamefont
  {Mattern}}, \bibinfo {author} {\bibfnamefont {G.~T.}\ \bibnamefont
  {Seidler}}, \bibinfo {author} {\bibfnamefont {G.}~\bibnamefont {Gregori}},
  \bibinfo {author} {\bibfnamefont {R.~W.}\ \bibnamefont {Falcone}}, \ and\
  \bibinfo {author} {\bibfnamefont {S.~H.}\ \bibnamefont {Glenzer}},\
  }\bibfield  {title} {\enquote {\bibinfo {title} {Observations of continuum
  depression in warm dense matter with x-ray thomson scattering},}\ }\href
  {\doibase 10.1103/PhysRevLett.112.145004} {\bibfield  {journal} {\bibinfo
  {journal} {Phys. Rev. Lett.}\ }\textbf {\bibinfo {volume} {112}},\ \bibinfo
  {pages} {145004} (\bibinfo {year} {2014})}\BibitemShut {NoStop}%
\bibitem [{\citenamefont {Plagemann}\ \emph {et~al.}(2015)\citenamefont
  {Plagemann}, \citenamefont {R\"uter}, \citenamefont {Bornath}, \citenamefont
  {Shihab}, \citenamefont {Desjarlais}, \citenamefont {Fortmann}, \citenamefont
  {Glenzer},\ and\ \citenamefont {Redmer}}]{PhysRevE.92.013103}%
  \BibitemOpen
  \bibfield  {author} {\bibinfo {author} {\bibfnamefont {K.-U.}\ \bibnamefont
  {Plagemann}}, \bibinfo {author} {\bibfnamefont {H.~R.}\ \bibnamefont
  {R\"uter}}, \bibinfo {author} {\bibfnamefont {T.}~\bibnamefont {Bornath}},
  \bibinfo {author} {\bibfnamefont {M.}~\bibnamefont {Shihab}}, \bibinfo
  {author} {\bibfnamefont {M.~P.}\ \bibnamefont {Desjarlais}}, \bibinfo
  {author} {\bibfnamefont {C.}~\bibnamefont {Fortmann}}, \bibinfo {author}
  {\bibfnamefont {S.~H.}\ \bibnamefont {Glenzer}}, \ and\ \bibinfo {author}
  {\bibfnamefont {R.}~\bibnamefont {Redmer}},\ }\bibfield  {title} {\enquote
  {\bibinfo {title} {Ab initio calculation of the ion feature in x-ray thomson
  scattering},}\ }\href {\doibase 10.1103/PhysRevE.92.013103} {\bibfield
  {journal} {\bibinfo  {journal} {Phys. Rev. E}\ }\textbf {\bibinfo {volume}
  {92}},\ \bibinfo {pages} {013103} (\bibinfo {year} {2015})}\BibitemShut
  {NoStop}%
\bibitem [{\citenamefont {Mo}\ \emph {et~al.}(2018)\citenamefont {Mo},
  \citenamefont {Fu}, \citenamefont {Kang}, \citenamefont {Zhang},\ and\
  \citenamefont {He}}]{PhysRevLett.120.205002}%
  \BibitemOpen
  \bibfield  {author} {\bibinfo {author} {\bibfnamefont {C.}~\bibnamefont
  {Mo}}, \bibinfo {author} {\bibfnamefont {Z.}~\bibnamefont {Fu}}, \bibinfo
  {author} {\bibfnamefont {W.}~\bibnamefont {Kang}}, \bibinfo {author}
  {\bibfnamefont {P.}~\bibnamefont {Zhang}}, \ and\ \bibinfo {author}
  {\bibfnamefont {X.~T.}\ \bibnamefont {He}},\ }\bibfield  {title} {\enquote
  {\bibinfo {title} {First-principles estimation of electronic temperature from
  x-ray thomson scattering spectrum of isochorically heated warm dense
  matter},}\ }\href {\doibase 10.1103/PhysRevLett.120.205002} {\bibfield
  {journal} {\bibinfo  {journal} {Phys. Rev. Lett.}\ }\textbf {\bibinfo
  {volume} {120}},\ \bibinfo {pages} {205002} (\bibinfo {year}
  {2018})}\BibitemShut {NoStop}%
\bibitem [{\citenamefont {Dornheim}\ \emph {et~al.}(2022)\citenamefont
  {Dornheim}, \citenamefont {Böhme}, \citenamefont {Kraus}, \citenamefont
  {Döppner}, \citenamefont {Preston}, \citenamefont {Moldabekov},\ and\
  \citenamefont {Vorberger}}]{Dornheim2022}%
  \BibitemOpen
  \bibfield  {author} {\bibinfo {author} {\bibfnamefont {T.}~\bibnamefont
  {Dornheim}}, \bibinfo {author} {\bibfnamefont {M.}~\bibnamefont {Böhme}},
  \bibinfo {author} {\bibfnamefont {D.}~\bibnamefont {Kraus}}, \bibinfo
  {author} {\bibfnamefont {T.}~\bibnamefont {Döppner}}, \bibinfo {author}
  {\bibfnamefont {T.~R.}\ \bibnamefont {Preston}}, \bibinfo {author}
  {\bibfnamefont {Z.~A.}\ \bibnamefont {Moldabekov}}, \ and\ \bibinfo {author}
  {\bibfnamefont {J.}~\bibnamefont {Vorberger}},\ }\bibfield  {title} {\enquote
  {\bibinfo {title} {Accurate temperature diagnostics for matter under extreme
  conditions},}\ }\href {\doibase 10.1038/s41467-022-35578-7} {\bibfield
  {journal} {\bibinfo  {journal} {Nature Communications}\ }\textbf {\bibinfo
  {volume} {13}},\ \bibinfo {pages} {7911} (\bibinfo {year}
  {2022})}\BibitemShut {NoStop}%
\bibitem [{\citenamefont {Zhang}\ \emph {et~al.}(2022)\citenamefont {Zhang},
  \citenamefont {Yuan}, \citenamefont {Zhang}, \citenamefont {Liu},
  \citenamefont {Fang}, \citenamefont {Zhang}, \citenamefont {Liu},
  \citenamefont {Zhao}, \citenamefont {Dong}, \citenamefont {Liu},
  \citenamefont {Dai}, \citenamefont {Gu}, \citenamefont {Li}, \citenamefont
  {Zheng}, \citenamefont {Zhong},\ and\ \citenamefont {Zhang}}]{Zhang2022}%
  \BibitemOpen
  \bibfield  {author} {\bibinfo {author} {\bibfnamefont {Z.}~\bibnamefont
  {Zhang}}, \bibinfo {author} {\bibfnamefont {X.-H.}\ \bibnamefont {Yuan}},
  \bibinfo {author} {\bibfnamefont {Y.-H.}\ \bibnamefont {Zhang}}, \bibinfo
  {author} {\bibfnamefont {H.}~\bibnamefont {Liu}}, \bibinfo {author}
  {\bibfnamefont {K.}~\bibnamefont {Fang}}, \bibinfo {author} {\bibfnamefont
  {C.-L.}\ \bibnamefont {Zhang}}, \bibinfo {author} {\bibfnamefont {Z.-D.}\
  \bibnamefont {Liu}}, \bibinfo {author} {\bibfnamefont {X.}~\bibnamefont
  {Zhao}}, \bibinfo {author} {\bibfnamefont {Q.-L.}\ \bibnamefont {Dong}},
  \bibinfo {author} {\bibfnamefont {G.-Y.}\ \bibnamefont {Liu}}, \bibinfo
  {author} {\bibfnamefont {Y.}~\bibnamefont {Dai}}, \bibinfo {author}
  {\bibfnamefont {H.-C.}\ \bibnamefont {Gu}}, \bibinfo {author} {\bibfnamefont
  {Y.-T.}\ \bibnamefont {Li}}, \bibinfo {author} {\bibfnamefont
  {J.}~\bibnamefont {Zheng}}, \bibinfo {author} {\bibfnamefont {J.-Y.}\
  \bibnamefont {Zhong}}, \ and\ \bibinfo {author} {\bibfnamefont
  {J.}~\bibnamefont {Zhang}},\ }\bibfield  {title} {\enquote {\bibinfo {title}
  {Efficient energy transition from kinetic to internal energy in supersonic
  collision of high-density plasma jets from conical implosions},}\ }\href
  {\doibase 10.7498/aps.71.20220361} {\bibfield  {journal} {\bibinfo  {journal}
  {Acta Physica Sinica}\ }\textbf {\bibinfo {volume} {71}},\ \bibinfo {pages}
  {155201--1--155201--8} (\bibinfo {year} {2022})}\BibitemShut {NoStop}%
\bibitem [{\citenamefont {Gregori}\ \emph {et~al.}(2003)\citenamefont
  {Gregori}, \citenamefont {Glenzer}, \citenamefont {Rozmus}, \citenamefont
  {Lee},\ and\ \citenamefont {Landen}}]{PhysRevE.67.026412}%
  \BibitemOpen
  \bibfield  {author} {\bibinfo {author} {\bibfnamefont {G.}~\bibnamefont
  {Gregori}}, \bibinfo {author} {\bibfnamefont {S.~H.}\ \bibnamefont
  {Glenzer}}, \bibinfo {author} {\bibfnamefont {W.}~\bibnamefont {Rozmus}},
  \bibinfo {author} {\bibfnamefont {R.~W.}\ \bibnamefont {Lee}}, \ and\
  \bibinfo {author} {\bibfnamefont {O.~L.}\ \bibnamefont {Landen}},\ }\bibfield
   {title} {\enquote {\bibinfo {title} {Theoretical model of x-ray scattering
  as a dense matter probe},}\ }\href {\doibase 10.1103/PhysRevE.67.026412}
  {\bibfield  {journal} {\bibinfo  {journal} {Phys. Rev. E}\ }\textbf {\bibinfo
  {volume} {67}},\ \bibinfo {pages} {026412} (\bibinfo {year}
  {2003})}\BibitemShut {NoStop}%
\bibitem [{\citenamefont {Mo}\ \emph {et~al.}(2020)\citenamefont {Mo},
  \citenamefont {Fu}, \citenamefont {Zhang}, \citenamefont {Kang},
  \citenamefont {Zhang},\ and\ \citenamefont {He}}]{PhysRevB.102.195127}%
  \BibitemOpen
  \bibfield  {author} {\bibinfo {author} {\bibfnamefont {C.}~\bibnamefont
  {Mo}}, \bibinfo {author} {\bibfnamefont {Z.-G.}\ \bibnamefont {Fu}}, \bibinfo
  {author} {\bibfnamefont {P.}~\bibnamefont {Zhang}}, \bibinfo {author}
  {\bibfnamefont {W.}~\bibnamefont {Kang}}, \bibinfo {author} {\bibfnamefont
  {W.}~\bibnamefont {Zhang}}, \ and\ \bibinfo {author} {\bibfnamefont {X.~T.}\
  \bibnamefont {He}},\ }\bibfield  {title} {\enquote {\bibinfo {title}
  {First-principles method for x-ray thomson scattering including both elastic
  and inelastic features in warm dense matter},}\ }\href {\doibase
  10.1103/PhysRevB.102.195127} {\bibfield  {journal} {\bibinfo  {journal}
  {Phys. Rev. B}\ }\textbf {\bibinfo {volume} {102}},\ \bibinfo {pages}
  {195127} (\bibinfo {year} {2020})}\BibitemShut {NoStop}%
\bibitem [{\citenamefont {Mattern}\ and\ \citenamefont
  {Seidler}(2013)}]{10.1063/1.4790659}%
  \BibitemOpen
  \bibfield  {author} {\bibinfo {author} {\bibfnamefont {B.~A.}\ \bibnamefont
  {Mattern}}\ and\ \bibinfo {author} {\bibfnamefont {G.~T.}\ \bibnamefont
  {Seidler}},\ }\bibfield  {title} {\enquote {\bibinfo {title} {{Theoretical
  treatments of the bound-free contribution and experimental best practice in
  X-ray Thomson scattering from warm dense matter}},}\ }\href {\doibase
  10.1063/1.4790659} {\bibfield  {journal} {\bibinfo  {journal} {Physics of
  Plasmas}\ }\textbf {\bibinfo {volume} {20}},\ \bibinfo {pages} {022706}
  (\bibinfo {year} {2013})}\BibitemShut {NoStop}%
\bibitem [{\citenamefont {Gawne}\ \emph {et~al.}(2024)\citenamefont {Gawne},
  \citenamefont {Bellenbaum}, \citenamefont {Fletcher}, \citenamefont {Appel},
  \citenamefont {Baehtz}, \citenamefont {Bouffetier}, \citenamefont
  {Brambrink}, \citenamefont {Brown}, \citenamefont {Cangi}, \citenamefont
  {Descamps}, \citenamefont {Goede}, \citenamefont {Hartley}, \citenamefont
  {Herbert}, \citenamefont {Hesselbach}, \citenamefont {Höppner},
  \citenamefont {Humphries}, \citenamefont {Konôpková}, \citenamefont
  {Laso~Garcia}, \citenamefont {Lindqvist}, \citenamefont {Lütgert},
  \citenamefont {MacDonald}, \citenamefont {Makita}, \citenamefont {Martin},
  \citenamefont {Mishchenko}, \citenamefont {Moldabekov}, \citenamefont
  {Nakatsutsumi}, \citenamefont {Naedler}, \citenamefont {Neumayer},
  \citenamefont {Pelka}, \citenamefont {Qu}, \citenamefont {Randolph},
  \citenamefont {Rips}, \citenamefont {Toncian}, \citenamefont {Vorberger},
  \citenamefont {Wollenweber}, \citenamefont {Zastrau}, \citenamefont {Kraus},
  \citenamefont {Preston},\ and\ \citenamefont {Dornheim}}]{10.1063/5.0222072}%
  \BibitemOpen
  \bibfield  {author} {\bibinfo {author} {\bibfnamefont {T.}~\bibnamefont
  {Gawne}}, \bibinfo {author} {\bibfnamefont {H.}~\bibnamefont {Bellenbaum}},
  \bibinfo {author} {\bibfnamefont {L.~B.}\ \bibnamefont {Fletcher}}, \bibinfo
  {author} {\bibfnamefont {K.}~\bibnamefont {Appel}}, \bibinfo {author}
  {\bibfnamefont {C.}~\bibnamefont {Baehtz}}, \bibinfo {author} {\bibfnamefont
  {V.}~\bibnamefont {Bouffetier}}, \bibinfo {author} {\bibfnamefont
  {E.}~\bibnamefont {Brambrink}}, \bibinfo {author} {\bibfnamefont
  {D.}~\bibnamefont {Brown}}, \bibinfo {author} {\bibfnamefont
  {A.}~\bibnamefont {Cangi}}, \bibinfo {author} {\bibfnamefont
  {A.}~\bibnamefont {Descamps}}, \bibinfo {author} {\bibfnamefont
  {S.}~\bibnamefont {Goede}}, \bibinfo {author} {\bibfnamefont {N.~J.}\
  \bibnamefont {Hartley}}, \bibinfo {author} {\bibfnamefont {M.-L.}\
  \bibnamefont {Herbert}}, \bibinfo {author} {\bibfnamefont {P.}~\bibnamefont
  {Hesselbach}}, \bibinfo {author} {\bibfnamefont {H.}~\bibnamefont
  {Höppner}}, \bibinfo {author} {\bibfnamefont {O.~S.}\ \bibnamefont
  {Humphries}}, \bibinfo {author} {\bibfnamefont {Z.}~\bibnamefont
  {Konôpková}}, \bibinfo {author} {\bibfnamefont {A.}~\bibnamefont
  {Laso~Garcia}}, \bibinfo {author} {\bibfnamefont {B.}~\bibnamefont
  {Lindqvist}}, \bibinfo {author} {\bibfnamefont {J.}~\bibnamefont {Lütgert}},
  \bibinfo {author} {\bibfnamefont {M.~J.}\ \bibnamefont {MacDonald}}, \bibinfo
  {author} {\bibfnamefont {M.}~\bibnamefont {Makita}}, \bibinfo {author}
  {\bibfnamefont {W.}~\bibnamefont {Martin}}, \bibinfo {author} {\bibfnamefont
  {M.}~\bibnamefont {Mishchenko}}, \bibinfo {author} {\bibfnamefont {Z.~A.}\
  \bibnamefont {Moldabekov}}, \bibinfo {author} {\bibfnamefont
  {M.}~\bibnamefont {Nakatsutsumi}}, \bibinfo {author} {\bibfnamefont {J.-P.}\
  \bibnamefont {Naedler}}, \bibinfo {author} {\bibfnamefont {P.}~\bibnamefont
  {Neumayer}}, \bibinfo {author} {\bibfnamefont {A.}~\bibnamefont {Pelka}},
  \bibinfo {author} {\bibfnamefont {C.}~\bibnamefont {Qu}}, \bibinfo {author}
  {\bibfnamefont {L.}~\bibnamefont {Randolph}}, \bibinfo {author}
  {\bibfnamefont {J.}~\bibnamefont {Rips}}, \bibinfo {author} {\bibfnamefont
  {T.}~\bibnamefont {Toncian}}, \bibinfo {author} {\bibfnamefont
  {J.}~\bibnamefont {Vorberger}}, \bibinfo {author} {\bibfnamefont
  {L.}~\bibnamefont {Wollenweber}}, \bibinfo {author} {\bibfnamefont
  {U.}~\bibnamefont {Zastrau}}, \bibinfo {author} {\bibfnamefont
  {D.}~\bibnamefont {Kraus}}, \bibinfo {author} {\bibfnamefont {T.~R.}\
  \bibnamefont {Preston}}, \ and\ \bibinfo {author} {\bibfnamefont
  {T.}~\bibnamefont {Dornheim}},\ }\bibfield  {title} {\enquote {\bibinfo
  {title} {Effects of mosaic crystal instrument functions on x-ray thomson
  scattering diagnostics},}\ }\href {\doibase 10.1063/5.0222072} {\bibfield
  {journal} {\bibinfo  {journal} {Journal of Applied Physics}\ }\textbf
  {\bibinfo {volume} {136}},\ \bibinfo {pages} {105902} (\bibinfo {year}
  {2024})}\BibitemShut {NoStop}%
\bibitem [{\citenamefont {Baczewski}\ \emph {et~al.}(2016)\citenamefont
  {Baczewski}, \citenamefont {Shulenburger}, \citenamefont {Desjarlais},
  \citenamefont {Hansen},\ and\ \citenamefont
  {Magyar}}]{PhysRevLett.116.115004}%
  \BibitemOpen
  \bibfield  {author} {\bibinfo {author} {\bibfnamefont {A.~D.}\ \bibnamefont
  {Baczewski}}, \bibinfo {author} {\bibfnamefont {L.}~\bibnamefont
  {Shulenburger}}, \bibinfo {author} {\bibfnamefont {M.~P.}\ \bibnamefont
  {Desjarlais}}, \bibinfo {author} {\bibfnamefont {S.~B.}\ \bibnamefont
  {Hansen}}, \ and\ \bibinfo {author} {\bibfnamefont {R.~J.}\ \bibnamefont
  {Magyar}},\ }\bibfield  {title} {\enquote {\bibinfo {title} {X-ray thomson
  scattering in warm dense matter without the chihara decomposition},}\ }\href
  {\doibase 10.1103/PhysRevLett.116.115004} {\bibfield  {journal} {\bibinfo
  {journal} {Phys. Rev. Lett.}\ }\textbf {\bibinfo {volume} {116}},\ \bibinfo
  {pages} {115004} (\bibinfo {year} {2016})}\BibitemShut {NoStop}%
\bibitem [{\citenamefont {Chihara}(2000)}]{Chihara2000}%
  \BibitemOpen
  \bibfield  {author} {\bibinfo {author} {\bibfnamefont {J.}~\bibnamefont
  {Chihara}},\ }\bibfield  {title} {\enquote {\bibinfo {title} {Interaction of
  photons with plasmas and liquid metals - photoabsorption and scattering},}\
  }\href {\doibase 10.1088/0953-8984/12/3/303} {\bibfield  {journal} {\bibinfo
  {journal} {Journal of Physics: Condensed Matter}\ }\textbf {\bibinfo {volume}
  {12}},\ \bibinfo {pages} {231} (\bibinfo {year} {2000})}\BibitemShut
  {NoStop}%
\bibitem [{\citenamefont {Fortmann}\ \emph {et~al.}(2009)\citenamefont
  {Fortmann}, \citenamefont {Thiele}, \citenamefont {Fäustlin}, \citenamefont
  {Bornath}, \citenamefont {Holst}, \citenamefont {Kraeft}, \citenamefont
  {Schwarz}, \citenamefont {Toleikis}, \citenamefont {Tschentscher},\ and\
  \citenamefont {Redmer}}]{FORTMANN2009208}%
  \BibitemOpen
  \bibfield  {author} {\bibinfo {author} {\bibfnamefont {C.}~\bibnamefont
  {Fortmann}}, \bibinfo {author} {\bibfnamefont {R.}~\bibnamefont {Thiele}},
  \bibinfo {author} {\bibfnamefont {R.}~\bibnamefont {Fäustlin}}, \bibinfo
  {author} {\bibfnamefont {T.}~\bibnamefont {Bornath}}, \bibinfo {author}
  {\bibfnamefont {B.}~\bibnamefont {Holst}}, \bibinfo {author} {\bibfnamefont
  {W.-D.}\ \bibnamefont {Kraeft}}, \bibinfo {author} {\bibfnamefont
  {V.}~\bibnamefont {Schwarz}}, \bibinfo {author} {\bibfnamefont
  {S.}~\bibnamefont {Toleikis}}, \bibinfo {author} {\bibfnamefont
  {T.}~\bibnamefont {Tschentscher}}, \ and\ \bibinfo {author} {\bibfnamefont
  {R.}~\bibnamefont {Redmer}},\ }\bibfield  {title} {\enquote {\bibinfo {title}
  {Thomson scattering in dense plasmas with density and temperature
  gradients},}\ }\href {\doibase https://doi.org/10.1016/j.hedp.2009.03.013}
  {\bibfield  {journal} {\bibinfo  {journal} {High Energy Density Physics}\
  }\textbf {\bibinfo {volume} {5}},\ \bibinfo {pages} {208--211} (\bibinfo
  {year} {2009})}\BibitemShut {NoStop}%
\end{thebibliography}%

\end{document}